\input harvmac
\overfullrule=0pt
\font\authorfont=cmcsc10 \ifx\answ\bigans\else scaled\magstep1\fi
{\divide\baselineskip by 4
\multiply\baselineskip by 3
\def\prenomat{\matrix{\hbox{hep-th/9607202}&\cr  
\qquad\hbox{SWAT/125}&\cr
\qquad\hbox{DTP/9672}&\cr}}
\Title{$\prenomat$}{\vbox{\centerline{Multi-Instanton Calculus}
\vskip2pt
\centerline{in $N=2$  Supersymmetric Gauge Theory}
\vskip6pt
\centerline{II.  Coupling to Matter}
}}
\centerline{\authorfont Nicholas Dorey}
\bigskip
\centerline{\sl Physics Department, University College of Swansea}
\centerline{\sl Swansea SA2$\,$8PP UK $\quad$ \tt  
n.dorey@swansea.ac.uk}
\bigskip
\centerline{\authorfont Valentin V. Khoze}
\bigskip
\centerline{\sl Department of Physics, Centre for Particle Theory, 
University of Durham}
\centerline{\sl Durham DH1$\,$3LE UK $\quad$ \tt  
valya.khoze@durham.ac.uk}
\bigskip
\centerline{and}
\bigskip
\centerline{\authorfont Michael P. Mattis}
\bigskip
\centerline{\sl Theoretical Division T-8, Los Alamos National  
Laboratory}
\centerline{\sl Los Alamos, NM 87545 USA$\quad$ \tt  
mattis@pion.lanl.gov}
\vskip .3in
\noindent
We further discuss the $N=2$ superinstantons in $SU(2)$ gauge theory,
obtained from the general
self-dual solutions of topological charge $n$ 
constructed by Atiyah, Drinfeld, Hitchin and Manin (ADHM).
We realize the $N=2$ supersymmetry
algebra as actions on the  superinstanton moduli.
This allows us to recast in concise superfield notation
our previously obtained expression for the exact classical interaction
between $n$ ADHM superinstantons mediated by the adjoint Higgs bosons,  
and
moreover, to incorporate $N_F$ flavors of hypermultiplets.
We perform explicit 1- and 2-instanton
checks of the Seiberg-Witten prepotentials for all $N_F$ and
arbitrary hypermultiplet masses. Our results for the low-energy
couplings are all in precise 
agreement with the predictions of Seiberg and Witten 
except for $N_F=4$,  where we find a finite
renormalization of the coupling which is absent in the proposed 
solution.
\vskip .2in
\vfil\break
}

\lref\CGOT{E. Corrigan, P. Goddard, H. Osborn and S. Templeton,
Nucl.~Phys.~B159 (1979) 469.}
\lref\Ohta{{Y. Ohta, \it Prepotentials of $N=2$ $SU(2)$ Yang-Mills
gauge theory coupled with a massive matter multiplet}, hep-th/9604051, and
 {\it Prepotentials of $N=2$ $SU(2)$ Yang-Mills
theories coupled with massive matter multiplets}, hep-th/9604059. }
\lref\Ito{K. Ito and S. K. Yang, {\it Picard-Fuchs equations and
prepotentials in $N=2$ supersymmetric QCD}, hep-th/9603073, to appear in
Proc. ``Frontiers of quantum field theory'', Toyonaka, Japan (1995).  }
\lref\Fucito{F. Fucito and G. Travaglini, 
{\it Instanton calculus and nonperturbative relations
 in $N=2$ supersymmetric gauge theories}, 
 ROM2F-96-32, hep-th/9605215.}
\lref\Osborntwo{H. Osborn,  Nucl. Phys. B140 (1978) 45.} 
\lref\Semi{A. Semikhatov, Phys. Lett. B120 (1982) 171.}
\lref\Sie{ W. Siegel, Phys Rev. (1995) D 52 1042}
\lref\Sadhm{A. Galperin and E. Sokatchev, Nucl. Phys. B460 (1995) 431;
S. J. Gates and L. Rana, Phys. Lett B345 (1995) 233; 
E. Witten, J. Geom. Phys. 15 (1995) 215.}
\lref\Gates{J. Gates, Nucl. Phys. B238 (1984) 349.}
\lref\dkmone{N. Dorey, V.V. Khoze and M.P. Mattis, \it Multi-instanton
calculus in $N=2$ supersymmetric gauge theory\rm, hep-th/9603136,
Phys.~Rev.~D (in press).}
\lref\dkmtwo{N. Dorey, V.V. Khoze and M.P. Mattis, 
\it Multi-instanton check of the relation between the prepotential  
${\cal F}$ 
and the modulus $u$ in $N=2$ SUSY Yang-Mills theory\rm,
hep-th/9606199.}
\lref\dkmthree{N. Dorey, V.V. Khoze and M.P. Mattis, 
\it A two-instanton test of the exact solution of $N=2$ supersymmetric  
QCD\rm,
hep-th/9607066.}
\lref\Aoyama{H. Aoyama, T. Harano, M. Sato and S.Wada,
\it   Multi-instanton calculus in N=2 supersymmetric QCD\rm,
hep-th/9607076.}   
\lref\Fayet{See for example P. Fayet and S. Ferrara, Phys.~Rep.~32  
(1977)
249, Sec.~2.2.}
\lref\Jack{I. Jack, Nucl. Phys. B174 (1980) 526.}
\lref\GSW{R. Grimm, M. Sohnius and J. Wess, Nucl. Phys. B133 (1978)  
275.}
\lref\BPST{A. Belavin, A. Polyakov, A. Schwartz and
Y. Tyupin, Phys. Lett. 59B (1975) 85.}
\lref\wessbagger{J. Wess and J. Bagger, {\it Supersymmetry and  
supergravity}, 
Princeton University Press, 1992.} 
\lref\Amati{A comprehensive review may be found in,
D. Amati, K. Konishi, Y. Meurice, G. Rossi and G. Veneziano,
Phys. Rep. 162 (1988) 169.}
\lref\Clark{T. E. Clark, O. Piguet and K. Sibold, Nucl. Phys. B143
(1978) 445.}
\lref\SWone{N. Seiberg and E. Witten, 
{\it Electric-magnetic duality, monopole
condensation, and confinement in $N=2$ supersymmetric Yang-Mills  
theory}, 
Nucl. Phys. B426 (1994) 19, (E) B430 (1994) 485  hep-th/9407087}
\lref\SWtwo{
N. Seiberg and E. Witten, 
{\it Monopoles, duality and chiral symmetry breaking
in $N=2$ supersymmetric QCD}, 
Nucl. Phys B431 (1994) 484 ,  hep-th/9408099}
\lref\KLTYone{A. Klemm, W. Lerche, S. Theisen and S. Yankielowicz 
{\it Simple singularities and $N=2$ supersymmetric Yang-Mills theory},  
Phys. Lett. B344 (1995) 169, hep-th/9411048}
\lref\AFone{P.C. Argyres and A.E. Faraggi, 
{\it The vacuum structure and spectrum of $N=2$ 
supersymmetric $SU(N)$ gauge theory},
 Phys. Rev. Lett. 74 (1995) 3931 , hep-th/9411057.}
\lref\KLTtwo{A. Klemm, W. Lerche and S. Theisen, 
{\it Nonperturbative effective actions of 
$N=2$ supersymmetric gauge theories}, 
CERN-TH/95-104, hep-th/9505150.}
\lref\DS{ M. Douglas and S. Shenker,
{\it Dynamics of $SU(N)$ supersymmetric gauge theory},
Nucl. Phys. B447 (1995) 271, hep-th/9503163; \hfil\break
U.H. Danielsson and B. Sundborg,
{\it The moduli space and monodromies of $N=2$ supersymmetric  
$SO(2R+1)$
Yang-Mills theory},
Phys. Lett. B358 (95) 273,  hep-th/9504102; \hfil\break 
A. Brandhuber and K. Landsteiner,
 {\it On the monodromies of $N=2$ supersymmetric $SO(2N)$}, 
Phys. Lett. B358 (1995) 73, hep-th/9507008; \hfil\break   
A. Hanany and Y. Oz,
{\it On the quantum moduli space of vacua of $N=2$ supersymmetric
$SU(N)$ gauge theories},
Nucl. Phys. B452 (1995) 283, hep-th/9505075;\hfil\break 
P. Argyres, M. Plesser and A. Shapere,
{\it The Coulomb phase of $N=2$ supersymmetric QCD},
 Phys. Rev. Lett. 75 (1995) 1699    hep-th/9505100; \hfil\break
P. Argyres and A. Shapere,
{\it The vacuum structure of N=2 super QCD with
 classical gauge groups},  RU-95-61,  hep-th/9505175; \hfil\break
A. Hanany,
{\it On the quantum moduli space of N=2
 supersymmetric gauge theories}, IASSNS-HEP-95/76,   hep-th/9505176. }
\lref\BIone{For reviews of recent progress in the field,  
see:\hfil\break
A. Bilal, {\it Duality in $N=2$ SUSY $SU(2)$ Yang-Mills theory:
A pedagogical introduction to the work of Seiberg and Witten}, 
LPTENS-95-53, hep-th/9601007; \hfil\break C. Gomez and R. Hernandez, 
{\it Electric-magnetic duality and effective field theories}, 
FTUAM 95/36,  hep-th/9510023.}
\lref\MAtone{
M. Matone, {\it Instantons and recursion relations in $N=2$ SUSY gauge  
theory}
 Phys. Lett. B357 (1995) 342,    hep-th/9506102.  }
\lref\FPone{ D. Finnell and P. Pouliot,
{\it Instanton calculations versus exact results in 4 dimensional 
SUSY gauge theories},
Nucl. Phys. B453 (95) 225, hep-th/9503115. }
\lref\ISone{ K.Ito and N. Sasakura,    
{\it One instanton calculations in $N=2$ supersymmetric $SU(N_c)$
Yang-Mills theory},
KEK-TH-470, hep-th/9602073.       }
\lref\FBone{ F. Ferrari and A. Bilal,    
{\it The strong coupling spectrum of the Seiberg-Witten theory},
LPTENS-96-16,  hep-th/9602082.}
\lref\Sone{ N. Seiberg, Phys. Lett. B206 (1988) 75. }
\lref\ADSone{ I. Affleck, M. Dine and N. Seiberg, Nucl. Phys. B241
(1984) 493; Nucl. Phys. B256 (1985) 557.  }
\lref\Affleck{I. Affleck, Nucl. Phys. B191 (1981) 429.}
\lref\Cone{ S. Cordes, Nucl. Phys. B273 (1986) 629.}
\lref\NSVZ{ V. A. Novikov, M. A. Shifman, A. I. Vainshtein and
V. I. Zakharov, Nucl Phys. B229 (1983) 394; Nucl. Phys. B229 (1983)
407; Nucl. Phys. B260 (1985) 157. }
\lref\FSone{ J. Fuchs and M. G. Schmidt, Z. Phys. C30 (1986) 161. }
\lref\FUone{ J. Fuchs, Nucl Phys B272 (1986) 677; Nucl Phys B 282
(1987) 437. }
\lref\ADHM{ M. F. Atiyah, V. G. Drinfeld, N. J. Hitchin and
Yu. I. Manin, Phys. Lett. A65 (1978) 185. }
\lref\ADHMtwo{ V. G. Drinfeld and Yu. I. Manin, Commun. Math. Phys. 
63 (1978) 177.} 
\lref\Oone{ H. Osborn, Ann. Phys. 135 (1981) 373. }
\lref\CGTone{ E. Corrigan, P. Goddard and S. Templeton,
Nucl. Phys. B151 (1979) 93; \hfil\break
   E. Corrigan, D. Fairlie, P. Goddard and S. Templeton,
    Nucl. Phys. B140 (1978) 31.}
\lref\CGone{ E. Corrigan and P. Goddard, {\it Some aspects of
instantons}, DAMTP 79/80, published in
the Proceedings of \it Geometrical and Topological Methods in Gauge 
Theories\rm,  Montreal 1979, eds. J.P. Harnad and S. Shnider,
Springer Lecture Notes in Physics 129 (Springer Verlag 1980).}
\lref\CWSone{ N. H. Christ, E. J. Weinberg and N. K. Stanton, Phys
Rev D18 (1978) 2013. }
\lref\GMOone{ P. Goddard, P. Mansfield and H. Osborn, Phys. Lett. B98 
(1981) 59. }
\lref\MAone{ P. Mansfield, Nucl. Phys. B186 (1981) 287. }
\lref\JNRone{ R. Jackiw, C. Nohl and C. Rebbi, Phys. Rev. D15 (1979)
1642.  }
\lref\Olive{C. Montonen and D. Olive, Phys. Lett. 72B (1977) 117.} 
\lref\oldyung{A. Yung, Nucl. Phys. B 344 (1990) 73.}
\lref\yung{A. Yung, {\it Instanton-induced effective
Lagrangian in the Seiberg-Witten model}, hep-th/9605096.}
\lref\tHooft{G. 't Hooft, Phys. Rev. D14 (1976) 3432; ibid.
D18 (1978) 2199.}
\lref\Derrick{G. H. Derrick,  J. Math.~Phys.~5 \rm (1964) 1252;
R. Hobart,  Proc.~Royal.~Soc. London  82 \rm (1963) 201.}

\def\frac#1#2{{ {#1}\over{#2}}}
\def\omegabar{\bar\omega}
\def\Kt{\tilde\K}
\def\bigL{{\bf L}}
\def\tot{{\rm tot}}
\def\Aone{{A^{\scriptscriptstyle(1)}}}
\def\Atwo{{A^{\scriptscriptstyle(2)}}}

\def\eff{{\rm eff}}
\def\inst{{\rm inst}}
\def\higgs{{\rm higgs}}
\def\K{{\cal K}}

\def\trtwo{\tr^{}_2\,}
\def\finv{f^{-1}}
\def\Ubar{\bar U}
\def\wbar{\bar w}

\def\abar{\bar a}
\def\bbar{\bar b}
\def\Deltabar{\bar\Delta}
\def\dalpha{{\dot\alpha}}
\def\dbeta{{\dot\beta}}
\def\dgamma{{\dot\gamma}}

\def\Im{{\rm Im}}

\def\sst{\scriptscriptstyle}

\def\vsd{v^{\sst\rm SD}}

\def\F{{\cal F}}
\def\P{{\cal P}}
\def\A{{\cal A}}
\def\susy{supersymmetry}
\def\sigmabar{\bar\sigma}

\def\cl{{\,\rm cl}}
\def\lambdabar{\bar\lambda}

\def\psibar{\bar\psi}
\def\sqrtwo{\sqrt{2}\,}

\def\thetabar{\bar\theta}
\def\Qbar{\bar Q}
\def\susic{supersymmetric}
\def\vhiggs{{\rm v}}
\def\vhiggsa{{{\cal A}_{\sst00}}}

\def\vbarhiggs{\bar{\rm v}}
\def\vhiggsbar{\bar{\rm v}}

\def\Abar{A^\dagger}

\def\C{{\cal C}}
\def\Cf{{\cal C}_{\sst\cal N}}

\def\zero{{\scriptscriptstyle(0)}}
\def\new{{\scriptscriptstyle\rm new}}

\def\uA{\,\lower 1.2ex\hbox{$\sim$}\mkern-13.5mu A}
\def\uX{\,\lower 1.2ex\hbox{$\sim$}\mkern-13.5mu X}
\def\uD{\,\lower 1.2ex\hbox{$\sim$}\mkern-13.5mu {\rm D}}

\def\uF{\,\lower 1.2ex\hbox{$\sim$}\mkern-13.5mu F}
\def\uW{\,\lower 1.2ex\hbox{$\sim$}\mkern-13.5mu W}
\def\uWbar{\,\lower 1.2ex\hbox{$\sim$}\mkern-13.5mu {\overline W}}
\def\uV{\,\lower 1.2ex\hbox{$\sim$}\mkern-13.5mu V}
\def\uv{\,\lower 1.0ex\hbox{$\scriptstyle\sim$}\mkern-11.0mu v}
\def\uPsi{\,\lower 1.2ex\hbox{$\sim$}\mkern-13.5mu \Psi}
\def\uPhi{\,\lower 1.2ex\hbox{$\sim$}\mkern-13.5mu \Phi}
\def\uchi{\,\lower 1.5ex\hbox{$\sim$}\mkern-13.5mu \chi}
\def\Psibar{\bar\Psi}
\def\uPsibar{\,\lower 1.2ex\hbox{$\sim$}\mkern-13.5mu \Psibar}
\def\upsi{\,\lower 1.5ex\hbox{$\sim$}\mkern-13.5mu \psi}
\def\psibar{\bar\psi}
\def\upsibar{\,\lower 1.5ex\hbox{$\sim$}\mkern-13.5mu \psibar}
\def\upsibarzero{\,\lower 1.5ex\hbox{$\sim$}\mkern-13.5mu  
\psibar^\zero}
\def\ulambda{\,\lower 1.2ex\hbox{$\sim$}\mkern-13.5mu \lambda}
\def\ulambdabar{\,\lower 1.2ex\hbox{$\sim$}\mkern-13.5mu \lambdabar}
\def\ulambdabarzero{\,\lower 1.2ex\hbox{$\sim$}\mkern-13.5mu  
\lambdabar^\zero}
\def\ulambdabarnew{\,\lower 1.2ex\hbox{$\sim$}\mkern-13.5mu  
\lambdabar^\new}
\def\D{{\cal D}}
\def\M{{\cal M}}
\def\N{{\cal N}}
\def\Dslash{\,\,{\raise.15ex\hbox{/}\mkern-12mu \D}}
\def\Dbarslash{\,\,{\raise.15ex\hbox{/}\mkern-12mu {\bar\D}}}
\def\delslash{\,\,{\raise.15ex\hbox{/}\mkern-9mu \partial}}
\def\delbarslash{\,\,{\raise.15ex\hbox{/}\mkern-9mu {\bar\partial}}}
\def\L{{\cal L}}
\def\hf{{\textstyle{1\over2}}}
\def\thf{{\textstyle{3\over2}}}
\def\quarter{{\textstyle{1\over4}}}
\def\sixteenth{{\textstyle{1\over16}}}

\def\xibar{\bar\xi}
\def\uAcl{\,\lower 1.2ex\hbox{$\sim$}\mkern-13.5mu A^{}_{\cl}}
\def\uAbarcl{\,\lower 1.2ex\hbox{$\sim$}\mkern-13.5mu A_{\cl}^\dagger}

\def\xione{\xi_1}
\def\xionebar{\bar\xi_1}
\def\xitwo{\xi_2}
\def\xitwobar{\bar\xi_2}

\def\Leff{\L_{\rm eff}}

\def\Atot{{\A_{\rm tot}}}
\def\Lambdatot{{\Lambda_{\rm tot}}}
\def\Qonebar{\bar Q_1}
\def\Qtwobar{\bar Q_2}
\newsec{Introduction}

\subsec{Recent background}

The low-energy dynamics of $N=2$ supersymmetric gauge theory in the
Coulomb phase is determined by a single holomorphic function: the
prepotential ${\cal F}$. In the case of $N=2$ supersymmetric
Yang-Mills (SYM) theory with gauge group $SU(2)$, an exact solution
for $\F$ has been obtained
by Seiberg and Witten \SWone. In Ref.~\SWtwo, 
these authors have generalized their analysis to include the coupling  
to 
$N_{F}$ flavors of matter hypermultiplets in the
fundamental representation of the gauge group. 
Their analysis relies on an elegant physical interpretation of the 
singularities of ${\cal F}$, as points at which the theory admits a
weakly-coupled dual description in terms of massless monopoles and  
dyons.

An important feature of the Seiberg-Witten analysis is that it  
comprises
a complete set of
predictions for all multi-instanton contributions to the long-distance
physics.  In principle, these predictions can  be compared with the
results of supersymmetric instanton calculus at weak coupling. 
Semiclassical instanton methods rely neither on duality, nor on any
subtle assumptions about the number or nature of the singularities
of $\F$ at strong coupling. As such, they provide 
independent tests of the proposed exact results of  
\refs{\SWone-\SWtwo},
and consequently,  of the electric-magnetic duality on which they are  
grounded.

This instanton program  has been carried out 
in the one- and two-instanton  sectors of $N=2$ SYM theory
in Refs.~\FPone\ and \dkmone,  respectively. (Another approach is
that of Ref.~\yung; also see Ref.~\ISone\
for higher gauge groups than $SU(2)$.) A new feature
 in the presence of massless matter hypermultiplets is that only even 
numbers of instantons contribute, due to an
anomalous discrete symmetry \SWtwo. Recently we have extended our
two-instanton analysis to this case as well \dkmthree, focusing on
the four-fermion vertex in the low-energy effective action. In a
parallel calculation,
the authors of Ref.~\Aoyama\ have extracted the two-instanton  
contribution
to  the expectation value of the quantum modulus
$u=\langle {\rm Tr} A^{2} \rangle$, with $A$ the adjoint Higgs field. 
The generalization to the case of massive hypermultiplets was also
briefly described in \dkmthree. So far, virtually
 all the instanton calculations described above have 
precisely confirmed the predictions of Seiberg and Witten. 
The sole exception has been a discrepancy \Aoyama\ in the
two-instanton contribution to $u$ in the model with $N_{F}=3$. 
In the following, we will also find an interesting discrepancy in the
case $N_{F}=4$. 

 In the absence of matter, it has also been possible to 
make some progress for  arbitrary instanton number $n$. The 
relevant field configurations are constrained supersymmetric
instantons based on the general 
solutions of the self-dual Yang-Mills equation obtained by 
Atiyah, Drinfeld, Hitchin and Manin (ADHM) \ADHM. In Ref.~\dkmone, 
we solved for the large- and short-distance behavior of all the  
component
fields in the self-dual background. This enabled us to construct
the exact classical interaction
between $n$ ADHM instantons mediated by the adjoint Higgs bosons,
both in the pure bosonic as well as  in the $N=2$ SYM theory. 

Unfortunately,  the problem of specifying the
multi-instanton measure for integration over the ADHM moduli remains
unsolved for $n>2$; this is the principal obstruction to an all-orders
check of the prepotential. Nevertheless, 
it is still possible to verify with our methods certain general
features of the proposed exact solution in the $n$-instanton sector. 
An example is the non-perturbative relation between the vacuum modulus  
$u$ and
the prepotential ${\cal F}$. While 
this relation was originally derived from 
the Seiberg-Witten solution by Matone \MAtone, 
it  turns out to be true on much broader grounds; 
in fact, it is built into the
instanton approach. This was shown in Ref.~\dkmtwo, which extends to  
all
$n$ the observation of Ref.~\Fucito.

The first goal of this paper is to generalize the multi-instanton 
SYM results of Ref.~\dkmone\  to allow for  $N_F$ fundamental  
hypermultiplets 
with arbitrary masses. We will 
construct the superinstanton action in this larger class of models,
extending to all $n$ the 2-instanton formula used in 
Refs.~\refs{\dkmthree,\Aoyama}. To accomplish this, we adopt a method 
which was originally developed in the early papers on supersymmetric  
instantons
by Novikov, Shifman, Vainshtein and Zakharov \NSVZ. As the relevant
field configurations obey equations of motion which are manifestly
supersymmetric, any non-vanishing action of 
the $N=2$ supersymmetry generators on a 
particular solution necessarily yields another solution. It follows
that the supersymmetry transformations of the fields are equivalent (up  
to a
gauge transformation) to certain transformations of 
the bosonic and fermionic collective coordinates 
of the superinstanton solution. Physically relevant quantities such as the
saddle-point action of the superinstanton must be constructed out of
\susic\ invariant combinations of the collective coordinates.   

An especially attractive feature of the ADHM
construction is that the various constraints on the parameters 
of the bosonic and fermionic fields are automatically supersymmetric. 
This means that the $N=2$ supersymmetry algebra can be realized  
directly as transformations of 
the highly over-complete (order $n^2$) set of collective
coordinates which appear explicitly
in the ADHM construction.\foot{This feature of the ADHM construction 
has been noticed by other authors \refs{\Semi\Sadhm-\Sie}.}
In fact, these parameters assemble naturally
into a single space-time-constant $N=2$ chiral superfield. 
This superfield notation 
systematizes the construction of supersymmetric invariant
combinations of the collective coordinates. Thus we demonstrate, in
retrospect, the invariance of the $N=2$ SYM superinstanton action
 obtained by component methods in \dkmone. The incorporation of 
$N_{F}$ massive hypermultiplets into this action is then
straightforward.

Subsequently, we apply these general formulae to some explicit 
calculations in  the one- and two-instanton sectors. 
These sections provide a more detailed account of the results
presented in our  recent letter \dkmthree. As stated there our results
agree with the predictions of Seiberg
and Witten for the four-antifermion correlator 
$\langle\bar{\lambda}(x_{1})\bar{\lambda}(x_{2})
\bar{\psi}(x_{3})\bar{\psi}(x_{4})\rangle$ which is proportional to 
$\partial^{4}{\cal F}/\partial {\rm v}^{4}$, where v is the classical
vacuum expectation value (VEV).
However, we now also extract the effective low-energy gauge coupling 
$\tau_\eff$ which involves the second derivative of the 
prepotential:\foot{The factor of $2$ on the right-hand side
of this equation is introduced in order to normalize $\tau_\cl$
 to the notation of \SWtwo, $\tau_{\rm cl}=\frac{8\pi
i}{g^{2}}+\frac{\theta}{\pi}$.
We perform no further rescalings
of the parameters of $N=2$ SYM theory when the hypermultiplets are 
added. See Ref.~\dkmone\ for a complete list of our conventions.}
\eqn\taudef{\tau_\eff\  = \ 2 \, 
\partial^{2}{\cal F}/\partial {\rm v}^{2}\ .}
The difference between these two quantities
as tests of the proposed exact solutions
is purely academic, except for the case $N_{F}=4$ which we now discuss.

\subsec{The case $N_F=4$.}

For $N_{F}=4$, both the perturbative \hbox{$\beta$-function} and the  
$U(1)_{R}$
anomaly vanish. This theory is parametrized by the
dimensionless gauge coupling $g^{2}$ and the vacuum angle $\theta$
which cannot be rotated away; these
 are combined to form a single 
complex parameter $\tau_{\rm cl}$.
In this model, Seiberg and Witten
propose an exact electric-magnetic duality 
which relates theories characterized by different values of
$\tau_{\rm cl}$. 
For this duality to hold it is necessary that the
massless $N_{F}=4$ theory be exactly conformally invariant. 
In other words, the conformal anomaly which vanishes to all orders
in perturbation 
theory must remain zero when non-perturbative effects are included. 
These authors also make the stronger assumption
 that the massless $N_{F}=4$ theory is 
classically exact, which means that the low-energy effective coupling
\taudef\  is simply equal to its classical counterpart:
\eqn\counterpart{\tau_{\rm eff}\ \equiv\  \tau_{\rm cl}\ .}
 In the massive $N_{F}=4$ theory 
 the low-energy correlators have, instead,  an infinite  expansion
in the dimensionless one-instanton factor $q$,
\eqn\qdeff{ q \ = \ \exp(i\pi\tau_{\rm cl}) \ . }

As mentioned above, certain general features of the multi-instanton 
contributions can be deduced from the general form of the
superinstanton action. In particular, we will obtain a representation 
for the prepotential itself as an integral over the superinstanton
moduli, generalizing to $N_F>0$ a result of Ref.~\dkmtwo. 
Although for $n>2$ the measure of integration is not known, its
only dependence on  ${\rm v}$ is through the superinstanton
action for which we have obtained an exact formula. By dimensional analysis,
each term in the multi-instanton series for $\F$ is seen to scale like
$\vhiggs^2$ when $N_F=4,$ so that the vanishing of the $\beta$-function
(which is related to $\F'''(\vhiggs)\,$) is essentially built into
the instanton approach.\foot{A caveat to this is that the integrals
over the superinstanton moduli must presumably be finite.} 

However, we should also
comment on a result we have obtained for the $N_{F}=4$  theory with
massless hypermultiplets
which appears to differ from the predictions of Seiberg and Witten. We  
have calculated the two-instanton contribution to the low-energy effective
coupling $\tau_{\rm eff}$ (the odd-instanton contributions being zero
as noted above).
We find that $\tau_{\rm eff}$ receives  finite corrections of
the form:  
\eqn\tauresult{
\tau_{\rm eff}\ =\ \tau_{\rm cl}\ +\ {i\over \pi}\sum_{n=2,4,6\ldots}\,
c_n\,q^{n}\ ,}
in contrast to the proposed classical exactness \counterpart.
Specifically we extract the dimensionless number $c_2$ in Sec.~8 below and
find that it is nonzero, $c_2\,=\,-7/(2^6\,3^5)\,.$ From our expression
for the prepotential
 we expect all the $c_n$ to be generically nonzero as well.
Note that the ``translation'' half of the modular group is preserved
by such a series: $\tau_\cl\rightarrow\tau_\cl+2$ implies
  $\tau_\eff\rightarrow\tau_\eff+2$. It is natural to conjecture that
the ``inversion'' half of the modular group is realized as well (albeit
in a more complicated way than for \counterpart) as some $SL(2,{\bf Z})$
transformation $\tau_\eff\,\rightarrow\,(a+b\tau_\eff)/(c+d\tau_\eff),$
with the value of $c_2$ providing a helpful clue.

Finally we should mention the case of  massless $N=4$ \susic\ gauge theory. 
In this case the instanton has   eight fermion zero modes which are protected 
by $N=4$ supersymmetry and cannot be lifted \Sone.
 It follows that 
instantons do not contribute to the prepotential and in particular corrections 
of the form \tauresult\ do not occur.

\subsec{The plan of this paper}

This paper is organized as follows. In Sec.~2, after a brief review
of our ADHM conventions from \dkmone, we implement the $N=2$ \susy\
algebra directly as an action on the over-complete set of bosonic
and fermionic ADHM parameters. The results of this exercise are
collected in Eq.~(2.28). In Sec.~3 we revisit the $N=2$ SYM multi-instanton
action from Ref.~\dkmone, and demonstrate that it is in fact 
a \susy\ invariant. To make this  manifest we assemble the
ADHM parameters into a single $N=2$ space-time-constant
``superfield,'' and recast the SYM action as an $N=2$ ``$F$-term.''
In Sec.~4 we show that the various ADHM bosonic and fermionic constraints
likewise assemble into a single $N=2$ supermultiplet.

Hypermultiplets are introduced starting in Sec.~5. We
incorporate them into the general multi-instanton SYM action using
invariance arguments. The final expression for the
 action, Eq.~(5.20), is discussed further in Sec.~6. 

In Sec.~7 we discuss the prepotential $\F$. General aspects of the prepotential
that emerge from the Seiberg-Witten formalism are reviewed in
Sec.~7.1. Alternatively, a formal representation, Eq.~(7.20),
of the prepotential
as an integral over the  multi-instanton supermoduli space is derived
in Sec.~7.2. 
Finally the explicit 1- and 2-instanton tests discussed above
of the proposed exact solutions 
are performed in Sec.~8. Numerical values of 
the 2-instanton contributions to the
prepotentials for $N_F=0,1,2,3,4$ are given in Eqs.~(8.21)-(8.22).

The paper also contains three Appendices. In particular, in Appendix C 
we explain an important difference in the numbers of 
fermion zero modes lifted by the 
VEV of an adjoint Higgs as opposed to a fundamental Higgs; in the
latter case the different topological sectors do not interfere.

\newsec{$N=2$ \susy\ algebra on the instanton moduli}

\subsec{ADHM preliminaries}

The basic object in the ADHM construction \ADHM\ of self-dual $SU(2)$  
gauge
fields of topological number $n$ is an $(n+1)\times n$  
quaternion-valued
matrix $\Delta_{\lambda l}(x)$, which is a linear function of the  
space-time
variable $x\,$:\foot{We use quaternionic notation $x=x_{\alpha\dalpha}
=x_n\sigma^n_{\alpha\dalpha},$ $\abar=\abar^{\dalpha\alpha}=\abar^n
\sigmabar_n^{\dalpha\alpha}$, $b=b_\alpha^{\ \beta},$ etc., where 
$\sigma^n$ and $\sigmabar^n$ are the spin matrices of Wess and Bagger
\wessbagger. 
See  the re-posted/published version of \dkmone\
for a self-contained introduction to the ADHM construction including a
full account of our conventions and a set of useful identities
used throughout the present paper.
We also set $g=1$ throughout, except, for clarity, in
the Yang-Mills instanton action $8\pi^2n/g^2.$
}
\eqn\Deltapostulate{\Delta_{\lambda l}
\ =\ a_{\lambda l}\ +\
b_{\lambda l}\,x\ ,\quad
0\le\lambda\le n\ ,\ \ 1\le l\le n\ .}
The gauge field $v_m(x)$ is then given by (displaying color indices)
\eqn\adhmansatz{v_m{}^\dalpha{}_\dbeta\ =\ 
\Ubar_\lambda^{\dalpha\alpha}\partial_m
U_{\lambda\,\alpha\dbeta}\ ,}
where the quaternion-valued vector $U_\lambda$ lives in the $\perp$  
space
of $\Delta\,$:
\eqna\perpspace
$$\eqalignno{\Deltabar_{l\lambda}\,U_{\lambda}
\ &=\ \Ubar_\lambda\,\Delta_{\lambda l}\ =\ 0\ ,&\perpspace a
\cr
\Ubar_\lambda U_\lambda\ &=\ 1\ .&\perpspace b}$$
It is easy to show that self-duality of the field strength $v_{mn}$ is
equivalent to the quaternionic condition
$\Deltabar^{\dbeta\beta}_{k\lambda}\,\Delta^{}_{\lambda l\,
\beta\dalpha}\, =\, (\finv)_{kl}\,\delta^\dbeta{}_\dalpha\,$; Taylor  
expanding
in $x$ then gives
\eqna\crucial
$$\eqalignno{\abar a\ &=\ (\abar a)^T\ \propto\  
\delta^\dbeta{}_\dalpha\
,&\crucial a
\cr
\bbar a\ &=\ (\bbar a)^T\ ,&\crucial b\cr
\bbar b\ &=\ (\bbar b)^T\  \propto\ \delta_\alpha{}^\beta\
,&\crucial c}$$
where the $\scriptstyle T$ stands for transpose in the ADHM indices
$(\lambda, l, \hbox{etc.})$ only.

In a \susic\ theory there is also the gaugino
\eqn\lambdazm{(\lambda_\alpha)^\dbeta{}_\dgamma\ =\ 
\Ubar^{\dbeta\gamma}\M_\gamma f\,\bbar\, U_{\alpha\dgamma}\ -\
\Ubar^\dbeta{}_\alpha \,bf\M^{\gamma T}U_{\gamma\dgamma}\ .}
We suppress ADHM indices but exhibit color (dotted) and Weyl
(undotted) indices for clarity. The condition that $\lambda$ be  
considered the superpartner of the self-dual gauge \hbox{field $v_m$} 
 is simply  that
it satisfy the 2-component Dirac equation in the ADHM background \NSVZ,
\hbox{$\Dbarslash^{\dalpha\alpha}\lambda_\alpha=0\,.$}
This, in turn, is equivalent to the following linear constraints
on the $(n+1)\times n$ constant Grassmann matrix $\M_\gamma$  
\CGTone:
\eqna\zmcons
$$\eqalignno{\abar^{\dalpha\gamma}\M_\gamma\ &=\ -\M^{\gamma T}a_\gamma
{}^\dalpha\ ,&\zmcons a
\cr
\bbar_\alpha{}^\gamma\M_\gamma\ &=\ \M^{\gamma T}b_{\gamma\alpha}\ .
&\zmcons b}$$
In the $N=2$ theory the fermion zero modes \lambdazm\ are reduplicated
by the Higgsino $\psi$ as well, to which we associate the matrix  
$\N_\gamma$;
in addition there is the adjoint Higgs field $A$ to be discussed
shortly.

Without loss of generality we can restrict to the following canonical  
forms 
for the $(n+1)\times n$ matrices $a_{\alpha\dalpha},$ $b_\alpha^{\  
\beta},$
$\M^\gamma$ and $\N^\gamma\,$:
\eqn\bcanonical{\eqalign{
a_{\alpha\dalpha}\ =\  
\pmatrix{w_{1\alpha\dalpha}&\cdots&w_{n\alpha\dalpha}
\cr{}&{}&{}\cr
{}&a'_{\alpha\dalpha}&{}\cr{}&{}&{}}\quad&,\qquad
b_\alpha^{\ \beta}\ =\ 
\pmatrix{0&\cdots&0 \cr \delta_\alpha{}^\beta & \cdots & 0 \cr
\vdots & \ddots & \vdots \cr 0 & \cdots &
\delta_\alpha{}^\beta}\ ,
\cr
\M^\gamma\ =\ \pmatrix{\mu_1^\gamma&\cdots&\mu_n^\gamma
\cr{}&{}&{}\cr
{}&\M^{\prime\gamma}&{}\cr{}&{}&{}}
\quad&,\qquad
\N^\gamma\ =\ \pmatrix{\nu_1^\gamma&\cdots&\nu_n^\gamma
\cr{}&{}&{}\cr
{}&\N^{\prime\gamma}&{}\cr{}&{}&{}}\ .}}
Thanks to this simple form for $b$, 
the constraint \crucial c is now automatically satisfied, while
 \crucial b and \zmcons b reduce to the symmetry
conditions on the $n\times n$ submatrices,
\eqn\symcond{a'\,=\,a^{\prime T}\quad,\quad
\M'\,=\,\M^{\prime T}\quad,\quad
\N'\,=\,\N^{\prime T}\ .}

Let us pause to count the number of superinstanton collective  
coordinates.
We expect there to be $8n$ bosonic degrees of freedom in the matrix $a$
($4n$ instanton positions, $n$ scale sizes, and $3n$ iso-orientations  
in
the far-separated limit), matched by $4n$ Grassmann degrees of  
freedom
in each of $\M$ and $\N.$ From \symcond, \zmcons a and \crucial a we  
see
that the fermionic count is correct; however there are far too many  
bosonic
variables, in fact, order $n^2$. These necessarily unphysical  
redundancies
reflect the existence of the remaining
$x$-independent $SU(2)\times O(n)$ symmetries which preserve all ADHM
constraints as well as the canonical form of $b$ given above:
\eqn\trans{\Delta\,\rightarrow\
\pmatrix{\Omega&0&\cdots&0\cr0&{}&{}&{}\cr\vdots&{}&R^T&{}\cr0&{}&{}&{}\cr}
\cdot\Delta\cdot R\ .}
Here $R\in O(n)$ and carries no $SU(2)$ indices; in contrast 
$\Omega_\alpha^{\ \beta}$ is a unit-normalized quaternion which acts on
these indices. While the $\Omega$ degrees of freedom merely
double-count the global gauge rotations, the action of the $O(n)$ poses
a bigger technical problem; in general it must be eliminated from the
path integral with a Faddeev-Popov prescription \refs{\Oone,\dkmone}.

\subsec{$N=2$ SUSY algebra}

We can now construct the $N=2$ \susy\ variation of the collective  
coordinate
matrix $a$. We start with the usual transformation law\foot{We follow  
the
\susy\ conventions of Appendix A of \dkmone, with the exception that
$v_m\rightarrow iv_m$ due to our conventional use of anti-Hermitian
ADHM gauge fields. The relation to the \susy\ parameters of \dkmthree\
is given by $\xi=\xi_1$ and $\xi'=-\xi_2.$} for the
gauge field under $\sum_{i=1,2}\,\xi_iQ_i+\xibar_i\Qbar_i\,$:
\eqn\deltav{\delta v_m\ =\ \xionebar\sigmabar_m\lambda
+\xitwobar\sigmabar_m\psi-\lambdabar\sigmabar_m\xione
-\psibar\sigmabar_m\xitwo\ .}
In the present case the first two terms on the right-hand side are  
obtained from 
Eq.~\lambdazm; the final two terms vanish, since the antifermions  
are
zero at the classical level (they are down by one power of the
coupling). Following  \NSVZ, the strategy is to trade
an \it active \rm transformation on the fields such as \deltav, for an
equivalent \it passive \rm transformation on the collective  
coordinates.
In order to do so, we must first understand \CGOT\ how to relate 
a variation $\delta U_\lambda$
(hence $\delta v_m$) to an underlying variation $\delta a$ of the  
collective
coordinate matrix, assuming  that we restrict attention to
 variations that preserve
the various ADHM conditions and constraints.
Varying \perpspace a gives $\delta\Ubar\,\Delta=-\Ubar\delta\Delta,$
or equivalently $\delta\Ubar\,(1-\P)=-\Ubar\delta\Delta f\Deltabar,$
where $\P$ is the usual ADHM projection operator
\eqn\Pdef{\P_{\lambda\kappa}\ =\ U_\lambda\Ubar_\kappa\ =\  
\delta_{\lambda
\kappa}-\Delta_{\lambda l}f_{lk}\Deltabar_{k\kappa}\ .}
By inspection, the general solution is 
\eqn\Utrans{\delta\Ubar=-\Ubar\delta\Delta f\Deltabar +\Sigma(x)\Ubar\  
.}
 Here $\Sigma(x)^\dalpha_{\ \dbeta}$ is arbitrary,
save for the condition \perpspace b which forces $0=\Sigma
+\bar\Sigma\equiv\trtwo \Sigma\,$;
consequently $\Sigma(x)$ is precisely an infinitesimal local $SU(2)$  
gauge
transformation. From \adhmansatz\ we obtain, finally \CGOT,
\eqn\passive{\eqalign{\delta v_m\ &=\ \delta\Ubar\partial_m U+\Ubar
\partial_m\delta U\cr&=\ \Ubar\delta a f\sigmabar_m\bbar U
-\Ubar\sigma_mbf\delta\abar U
-\D_m \Sigma\ .}}
In the last rewrite we have used \perpspace a, \Deltapostulate, and
an integration by parts; we have also set $\delta\Delta=\delta a$ as we  
are
holding $b$ fixed as per \bcanonical. Comparing the active  
transformation
\deltav\ and \lambdazm\ with the passive transformation \passive, we
extract the simple rule
\eqn\susyona{\delta a_{\alpha\dalpha}\ =\ \xibar_{1\dalpha}\M_\alpha
+\xibar_{2\dalpha}\N_\alpha}
Notice that the local gauge transformation
 represented by the last 
term of \passive\ proved unnecessary; one can set $\Sigma=0$.

Next we turn to the subtler case of the
fermions, whose active \susy\ transformation is given by:
\eqna\activef
\eqna\eulermatter
$$\eqalignno{\delta\lambda\ &=\ 
i\sqrtwo \xitwobar\Dbarslash A
-i\xione\sigma^{mn}v_{mn}\ ,&\activef a
\cr
\delta\psi\ &=\ -i\sqrtwo\xionebar\Dbarslash A
-i\xitwo\sigma^{mn}v_{mn}\ .
&\activef b}$$
Here the adjoint Higgs component $A$ of the superinstanton
is defined by the Euler-Lagrange equation\foot{In both
\activef{} and \eulermatter{}, and elsewhere in this paper,
we ignore the auxiliary fields $F$ and $D$
which only turn on at a higher order in the coupling.}
$$\eqalignno{\D^2 A\ &=\ \sqrtwo i\,[\,\lambda\,,\psi\,]\  
.&\eulermatter{}}$$
(In contrast, the antiboson $A^\dagger$ obeys the homogeneous equation
\eqn\Adageqn{\D^2 A^\dagger\ =\ 0}
when $N_F=0$; the superinstanton breaks the conjugation symmetry  
between 
$A$ and $A^\dagger$.) The construction of the
solution of \eulermatter{} for general $n$ is one of the principal
results \hbox{of \dkmone}. In brief, the answer has the additive form 
$A=\Aone+\Atwo$, where
\eqn\Aonedef{i\,\Aone^\dalpha_{\ \dbeta}\ =\
{1\over2\sqrtwo}\,\Ubar^{\dalpha\alpha}\big(\N_\alpha f\M^{\beta T}
-\M_\alpha f\N^{\beta T}\big)U_{\beta\dbeta}\ ,}
and 
$i\,\Atwo^\dalpha_{\ \dbeta} =
\Ubar^{\dalpha\alpha}\,\A_\alpha^{\ \beta}\,U_{\beta\dbeta} \ ,$
with $\A$  a block-diagonal constant matrix,
\eqn\blockdiag{\A_\alpha^{\ \beta}\ =\
\pmatrix{\vhiggsa_\alpha^{\ \beta}&0&\cdots&0 \cr 0&{}&{}&{}\cr
\vdots&{}&\Atot\,\delta_\alpha^{\ \beta}&{}\cr 0&{}&{}&{}}\ .}
$\vhiggsa$ is related in a trivial way to the VEV (which we point in
the $\tau^3$ direction),
\def\barvhiggsa{\bar{\cal A}_{\sst00}}
\eqn\vevbcagain{{\vhiggsa}_\alpha{}^\beta
\ =\ \textstyle{i\over2}\,\vhiggs\,\tau^3{}_\alpha{}^\beta\ ,\qquad
\barvhiggsa{}_\alpha{}^\beta
\ =\ -\textstyle{i\over2}\,\vbarhiggs\,\tau^3{}_\alpha{}^\beta\ ,
}
while the $n\times n$ antisymmetric matrix $\Atot$ is defined as the
solution to an inhomogeneous linear matrix equation \dkmone, namely
Eq.~(A.1) in  Appendix A below. (In the language of Sec.~7 of \dkmone,  
$\Atot$
is the sum
\eqn\Atotsum{\Atot\ =\ \A'+\A'_f}
where $\A'$ is purely bosonic while $\A'_f$ 
is a fermion bilinear.)

As above we need to equate the active transformation
\activef{a} with the \rm passive \rm transformation
derived from \lambdazm:
\eqn\passivef{\delta\lambda\ =\ 
\Ubar\delta\M f\bbar U+
\delta\Ubar\M f\bbar U+
\Ubar\M f\bbar\delta U+
\Ubar\M \delta f\bbar U\ -\ \hbox{H.c.}}
The first term on the right-hand side contains the unknown $\delta\M$
that we wish to determine; the second and third terms are already 
fixed by \Utrans\ and \susyona; the fourth term, too, is a known
entity, since 
\eqn\entity{\delta f\ =\ -f\delta(\Deltabar\Delta)f\ =\
-f(\delta\abar\Delta
+\Deltabar\delta a)f\ .}
 A lengthy but straightforward calculation yields
a welcome simplification: the second, third and fourth terms, taken
together, cancel precisely against the piece 
$i\sqrtwo \xitwobar\Dbarslash \Aone$ from \Aonedef\ 
that enters the right-hand side of \activef a.
Equating what remains gives the defining condition for $\delta\M\,$:
\eqn\deltaMeqn{\eqalign{
\Ubar^{\dbeta\gamma}\delta\M_\gamma f\,\bbar\, U_{\alpha\dgamma}\ -\
\hbox{H.c.}
\ &=\ 
i\sqrtwo \xitwobar\Dbarslash \Atwo-i\xi_1\sigma^{mn}v_{mn}
\cr
 &=\ 
\Ubar^{\dbeta\gamma}\big(-4ib\xi_{1\gamma}-2\sqrtwo \C_{\gamma\dalpha}
\xibar_2^\dalpha\big)f\bbar U_{\alpha\dgamma}
\ -\
\hbox{H.c.}}}
The final rewrite makes use of the 
well-known form of the ADHM field strength,
\eqn\vmnagain{v_{mn}{}^\dalpha{}_\dbeta\ =\ \big(
v_{mn}{}^\dalpha{}_\dbeta\big)^{\rm\sst dual}\ =\ 
4\Ubar^{\dalpha\alpha}
\,b\sigma_{mn\,\alpha}{}^\beta
\,f\,\bbar U_{\beta\dbeta}\ ,}
as well as identities (7.8) and (C1) from \dkmone.
The $(n+1)\times n$ quaternion-valued constant matrix $\C$ is defined  
as
\eqn\Cdef{\C\ =\
\pmatrix{\vhiggsa w_1-w_k\Atot{}_{k1}&\cdots&\vhiggsa  
w_n-w_k\Atot{}_{kn}
\cr
{}&{}&{}\cr
{}&\big[\,\Atot\,,\,a'\,]&{}\cr
{}&{}&{}  }\ .}
It follows that
\eqna\delMdef
$$\eqalignno{\delta\M_\gamma\ &=\ 
-4ib\xi_{1\gamma}-2\sqrtwo\C_{\gamma\dalpha}
\xibar_2^\dalpha
&\delMdef a}$$
and likewise
$$\eqalignno{\delta\N_\gamma\ &=\ 
-4ib\xi_{2\gamma}+2\sqrtwo \C_{\gamma\dalpha}
\xibar_1^\dalpha \ .
&\delMdef b}$$

The final ingredient needed is the $N=2$ transformation law for $\Atot$  
itself.
As shown in  Appendix A, it is a singlet:
$\delta\Atot = 0.$ This equation together with
Eqs.~\susyona\ and \delMdef{}  are the sought-after realization
of the $N=2$ \susy\ algebra on the collective coordinates of the
ADHM superinstanton. When hypermultiplets are included, these equations  
are
supplemented by Eq.~(5.18) below, where $\K$ and $\Kt$ are the  
Grassmann
collective coordinates associated with the fundamental fermions.
 For ease of reference we assemble them all here:
\eqna\susyalgebra
$$\eqalignno{
\delta a_{\alpha\dalpha}\ &=\ \xibar_{1\dalpha}\M_\alpha
+\xibar_{2\dalpha}\N_\alpha \qquad&\susyalgebra a \cr
\delta\M_\gamma\ &=\ -4ib\xi_{1\gamma}-2\sqrtwo\C_{\gamma\dalpha}
\xibar_2^\dalpha  \qquad&\susyalgebra b \cr
\delta\N_\gamma\ &=\ -4ib\xi_{2\gamma}+2\sqrtwo \C_{\gamma\dalpha}
\xibar_1^\dalpha \qquad&\susyalgebra c \cr
\delta\Atot\ &=\ 0\ \qquad&\susyalgebra d \cr
 \delta\K_i\ &=\  0 \qquad&\susyalgebra e \cr
\delta\Kt_i\ &=\ 0\qquad &\susyalgebra f }$$

The careful reader will notice, however, that
the $N=2$ algebra is not precisely obeyed by the above. For instance,
the anticommutator $\{\Qonebar,\Qtwobar\}$, rather than vanishing when
acting on $a,$ $\M$ or $\N$, gives a residual symmetry transformation
of the form \trans. (This is analogous to naive realizations of \susy\
that fail to commute with Wess-Zumino gauge fixing, for example.)
For present purposes this poses no problem,
as  we are always ultimately concerned with singlets under \trans;
otherwise one would have to covariantize the \susy\ transformations
with respect to \trans\ in the standard way.

\newsec{Multi-instanton action in pure $N=2$ \susic\ gauge theory}

Although, as we saw in the 
previous section, the superinstanton transforms under supersymmetry, its 
saddle-point action must be invariant. For a single instanton, in the
presence of a Higgs (fundamental or adjoint), the
bosonic part of the action is proportional to $|\vhiggs|^{2}\rho^2$. 
In $N=1$ models, such as those 
considered in \refs{\NSVZ,\FSone}, the squared instanton scale-size $\rho^{2}$ 
is augmented in the action by a fermion bilinear term to form a \susic\
  invariant combination  $\rho^{2}_{\rm inv}$.
 We now check  that the same property holds for
the action in $N=2$ \susic\ Yang-Mills theory, for arbitrary
topological number $n$. In this case the action is given by \dkmone
\def\Lambdabar{\bar\Lambda}
\def\vhiggsabar{\barvhiggsa}
\eqn\sinstfinal{S^0_{\rm inst}\ =\ 
{8n\pi^2\over g^2}\ +\ 16\pi^2|\vhiggsa|^2\sum_k|w_k|^2\ 
-\ 8\pi^2\Lambdabar_{lk}\A_{\tot\, kl}
+\  
4\sqrtwo\pi^2\,\mu_k^\alpha\,\barvhiggsa{}_\alpha{}^\beta\,\nu_{k\beta}
\ ,}
where $\Lambdabar$ is the $n\times n$ scalar-valued antisymmetric  
matrix
\eqn\Lambdabardef{\Lambdabar_{lk}\ =\ \wbar_l\vhiggsabar w_k
-\wbar_k\vhiggsabar w_l\ ,\qquad
\Lambda_{lk}\ =\ \wbar_l\vhiggsa w_k
-\wbar_k\vhiggsa w_l\ .}
The \susic\ invariance of this expression under the transformations
\susyalgebra{a\hbox{-}d}\ is immediate: the second and
third terms on the right-hand side give, respectively,
$-16\pi^2|\vhiggsa|^2(\mu_kw_k\xibar_1+\nu_kw_k\xibar_2)$ and
$-16\pi^2(\xibar_2\wbar_l\vhiggsabar\nu_k-\mu_l\vhiggsabar w_k\xibar_1)
\A_{\tot\, kl}$ which are canceled precisely by the variation of the
last term.

Despite the simplicity of this last calculation, it is illuminating to 
reformulate the action \sinstfinal\ in a more concise form in which
the \susy\ is manifest. To this end, we promote the ADHM collective
coordinate matrix $a$ to a space-time-constant
``superfield'' $a(\thetabar_i)$ in an
obvious way:\foot{From now on we ignore the action of the $Q_i$ which
act in a trivial way, and focus exclusively on the $\Qbar_i$. For
a related construction in a VEVless model, see \Semi.}
\eqn\apromote{\eqalign{a_{\alpha\dalpha}
\ &\longrightarrow\ a_{\alpha\dalpha}(\thetabar_i)\ =\ 
e^{\thetabar_2\Qbar_2}\times e^{\thetabar_1\Qbar_1}\times  
a_{\alpha\dalpha}
\cr&=\ 
a_{\alpha\dalpha}+\thetabar_{1\dalpha}\M_\alpha
+\thetabar_{2\dalpha}\N_\alpha
+2\sqrtwo\C_{\alpha\dbeta}\thetabar_2^\dbeta\thetabar_{1\dalpha}^{}
+\sqrtwo\thetabar_{1\dalpha}\thetabar_2^{\,2}\Cf{}_\alpha}}
where the Grassmann matrix $\Cf$ is defined in analogy with $\C$,
\eqn\Cfdef{\Cf\ =\
\pmatrix{\vhiggsa \nu_1-\nu_k\Atot{}_{k1}&\cdots&\vhiggsa 
\nu_n-\nu_k\Atot{}_{kn}
\cr
{}&{}&{}\cr
{}&\big[\,\Atot\,,\,\N'\,]&{}\cr
{}&{}&{}  }\ .}

\def\Pinfty{\P_{\sst\infty}}
A short calculation making use of the defining equation (A.1) for  
$\Atot$ gives
the desired rewrite of the action \sinstfinal\ as a manifestly \susic\
 $N=2$ ``$F$-term''$\,$:
\eqn\Scompact{S^0_{\rm inst}\ =\ {8n\pi^2\over g^2}\ 
-\ \pi^2\,\Tr\,\abar(\thetabar)\big(\Pinfty+1\big)a(\thetabar)\
\Big|_{\thetabar_1^2\thetabar_2^2}}
Here the capitalized `Tr' indicates a trace over both ADHM and $SU(2)$
indices, $\Tr=\Tr_n\circ\trtwo,$ and $\Pinfty$ is the  
$(n+1)\times(n+1)$ matrix
\eqn\Pinftydef{\Pinfty\ =\ \lim_{r\rightarrow\infty}\P\ =\ 1-b\bbar
\ =\ \delta_{\lambda0}\delta_{\kappa0}\ .}
Note the following:

1. The intermediate expression \apromote\ is not symmetric in  
$\thetabar_1$
and $\thetabar_2$. This merely reflects the point noted earlier,
that we have only realized the $N=2$ algebra up to transformations
of the type \trans; therefore
$\Qbar_1$ and $\Qbar_2$ do not actually anticommute. Nevertheless
the final expression \Scompact\ is a singlet under \trans, so
this poses no problems.

2. The purely bosonic part of $S_\inst^0$ in Eq.~\Scompact\ 
may be viewed in two
ostensibly different ways. On the one hand, in the above construction, 
it comes entirely from
the square of the fourth term on the right-hand side of \apromote.
On the other hand, we also know \dkmone\ that it
comes entirely from the Higgs kinetic energy term
in the component Lagrangian (where only the bosonic part of $A$ is
taken). To reconcile these two statements, note
 that the bosonic part of $\C$ from Eq.~\Cdef\
is precisely Eq.~(C.1) in \dkmone.
The bosonic action therefore corresponds to the expression (B.5) 
of \dkmone\ for
the overlap of two vector zero modes $\D_nA$, which is indeed
the Higgs kinetic
energy (see Appendices B and C of \dkmone\ for details). The form of
\Scompact\ was therefore inevitable.

\newsec{Supersymmetric reformulation of the constraint equations}

The component fields of the $N=2$ superinstanton, and their
respective moduli, follow a suggestive
 pattern:

(i) The gauge field $v_m$ obeys a nonlinear homogeneous
differential equation (the Yang-Mills equation). The associated
collective coordinates $a$ obey a nonlinear homogeneous constraint,
\crucial a. 
This condition  imposes $\thf n(n-1)$
constraints on the upper-triangular traceless quaternionic elements
of $\abar a.$

(ii) The fermions $\lambda$ and $\psi$ obey linear homogeneous
differential equations (the covariant Dirac equation). Their associated
moduli $\M$ and $\N$ obey the linear homogeneous constraint \zmcons a.
This imposes $n(n-1)$ conditions on each of $\M$ and
$\N.$

(iii) Finally the Higgs field $A$ is the solution to an inhomogeneous  
linear
differential equation (the covariant Klein-Gordon equation with a  
Yukawa
source term). Correspondingly, the matrix 
$\Atot$ satisfies an inhomogeneous linear ``constraint equation,''  
namely 
Eq.~(A.1) below. This equation  determines the $\hf n(n-1)$
scalar degrees of freedom in the $n\times n$ antisymmetric matrix  
$\Atot.$

Notice that the total number of bosonic and fermionic constraints
are each \hbox{$2n(n-1)$.}  This balancing between bosonic and fermionic  
degrees of 
freedom suggests that the set of constraints \crucial a, \zmcons a and  
(A.1)
might naturally be combined into an $N=2$  ``supermultiplet'' of  
constraints.
Here we show that this is in fact the case.

In light of the ``superfield'' $a(\thetabar)$ constructed above, the  
obvious
ansatz for this supermultiplet of constraints is to introduce  
$\thetabar$
dependence into
 the original ADHM condition \crucial a:
\eqn\superadhm{\abar(\thetabar) a(\thetabar)\ =\ \big(\abar(\thetabar)
 a(\thetabar)\big)^T\ \propto\ \delta^\dbeta{}_\dalpha\ .}
(Note that \crucial b is automatically satisfied for $a\rightarrow
a(\thetabar)$ thanks to the canonical choices \bcanonical-\symcond.)
The first few terms in the Taylor expansion of \superadhm\ look
promising (see Fig.~1): the bosonic component is just \crucial a  
itself, while
the $\thetabar_1$ and $\thetabar_2$ components indeed reproduce
the zero-mode condition \zmcons a for $\M$ and $\N$, respectively.

\vbox{
\vskip .3in
$$\matrix{{}&v_m&{}\cr\lambda&{}&\psi\cr{}&A&{}\cr}
\qquad\Longleftrightarrow\qquad
\matrix{{}&1&{}\cr\thetabar_1&{}&\thetabar_2\cr{}&\thetabar_1\thetabar_2 
&{}\cr}
$$
\centerline{$\hbox{\tenrm 
Fig.~1. The $N=2$ supermultiplet, and the corresponding}$}
\centerline{$\qquad\qquad\qquad$ elements of the super-ADHM constraints 
\superadhm.
$\qquad\qquad\qquad$
}
\vskip .3in
}

Less obvious is the $\thetabar_1\times\thetabar_2$ component of   
\superadhm,
which we  rewrite  as the triplet of conditions
\eqn\triplet{\trtwo\,\tau^k\,\abar(\thetabar)a(\thetabar)\ 
=\ 0\ ,\qquad k=1,2,3}
where $\tau^k$ is a Pauli matrix. Extracting the $\thetabar_{2\dbeta}
\thetabar^\dalpha_1$ component of \triplet\ after some index
rearrangement gives
\def\Cbar{\bar{\cal C}}
\eqn\crossterm{0\ =\ \tau^{k\dbeta}{}_\dalpha\,\Lambda_f
+\tau^k_{\dgamma\dalpha}\big(\,(\abar\C)^{\dgamma\dbeta}
+(\Cbar a)^{\dbeta\dgamma}\,\big)\ ,}
where in the notation of \dkmone,
\eqn\newmatdef{ \Lambda_f\ =\ -\Lambda_f^T\ =\ 
{1\over2\sqrtwo}\,
\big(\,\M^{\beta T}\N_\beta-\N^{\beta T}\M_\beta\,\big)\ . }
This equation is analyzed as follows. Tracing on color indices tells us  
that
$\abar\C+\Cbar a \propto \delta^\dalpha{}_\dbeta.$
So we plug 
\eqn\color{(\Cbar a)^\dalpha_{\ \dbeta}\ =\ X\delta^\dalpha_{\ \dbeta}
-(\abar\C)^\dalpha_{\ \dbeta}}
 into \crossterm, the $n\times n$ matrix $X$ being the unknown,
and deduce $X=\Lambda_f+\trtwo\abar\C$. Equation \color\ then becomes
\eqn\newcolor{\Lambda_f\ =\ \Cbar a+(\abar\C-\trtwo\abar\C)
\ =\ \Cbar a-(\Cbar a)^T\ .}
Up to this point the manipulations have been valid for 
arbitrary $\C.$ But if one substitutes the explicit 
expression \Cdef\ for $\C$ in terms of $\Atot,$
Eq.~\newcolor\ does in fact become the
defining linear equation (A.1) for $\Atot$, expressed in especially
concise form. With \Cdef\ and \crucial a
one also confirms that 
$\abar\C+\Cbar a$ is pure trace in the $SU(2)$ space; thus all tensor
components of \crossterm\ have properly been accounted for.
The remaining $\theta$ components of \superadhm\ turn out to 
be ``auxiliary'' as they contain
no new information. Some are satisfied trivially, while others
boil down to the earlier relations \crucial a, \zmcons a, or (A.1).

\newsec{Multi-instanton action in $N=2$ \susic\ QCD}
\def\Ahyp{{{\cal A}_{\rm hyp}}}
\def\hyp{{\rm hyp}}
\def\mass{{\rm mass}}
\def\Lambdahyp{{\Lambda_\hyp}}
Following \SWtwo\ we now turn our attention to the richer class of
models in which the $N=2$ \susic\ Yang-Mills action is augmented by
 $N_{F}$ matter hypermultiplets which transform in the fundamental
representation of $SU(2)$. Each $N=2$ hypermultiplet corresponds to a  
pair of
$N=1$ chiral multiplets, $Q_{i}$ and $\tilde{Q}_{i}$ where 
$i=1,2,\cdots, N_{F}$, which contain scalar quarks (squarks) $q_{i}$  
and
$\tilde{q}_{i}$ respectively and fermionic partners 
$\chi_{i}$ and $\tilde{\chi}_{i}$. 
We will restrict our attention to the Coulomb branch of
the theory where the squarks do not acquire a VEV.  
In the $N=1$ language, the matter fields couple to the gauge multiplet
via a superpotential, 
\eqn\superp{{\rm W}\ =\ \sum_{i=1}^{N_{F}} \sqrt{2}\tilde{Q}_{i}\Phi  
Q_{i} +
m_{i}\tilde{Q}_{i}Q_{i}}
suppressing color indices. The second term is an $N=2$ invariant mass  
term.

As reviewed above, the component fields  of the superinstanton which
reside in the adjoint representation of $SU(2)$  have the generic
form $\Ubar XU.$  Similarly, those in the fundamental representation
have the  structure $\Ubar X.$ The solution of the coupled
Euler-Lagrange equations for each of these fields is simplified by the
use of the differentiation identities in the ADHM background,
\eqna\diffids
$$\eqalignno{\D_n(\Ubar X)\ &=\ -\Ubar\partial_n\Delta f\Deltabar X
+\Ubar\partial_n X &\diffids a
\cr
\D^2(\Ubar X)\ &=\ \Ubar\partial^2X-2\Ubar b\sigma^nf
\Deltabar\partial_nX+4\Ubar bf\bbar X &\diffids b
\cr
\D_n(\Ubar XU)\  &=\ 
-\Ubar\partial_n\Delta f\Deltabar X U-\Ubar X\Delta  
f\partial_n\Deltabar\,U
+\Ubar\partial_n X\, U &\diffids c
\cr
\D^2(\Ubar XU)\ &=\ 
4\Ubar\,\big\{\,bf\,\bbar\,,\,X\,\big\}\,U
-4\Ubar b f\cdot\trtwo\Deltabar X\Delta\cdot f\,\bbar U
\cr&+\Ubar\,\partial^2X\,U
-2\Ubar\, bf\sigma_n\Deltabar\,\partial^n X\,U
-2\Ubar\partial^n X\Delta\sigmabar_nf\,\bbar U&\diffids d}$$

As in \dkmone, the construction of the short-distance
superinstanton starts with the
fermion zero modes in the ADHM background, 
then proceeds to the Higgs bosons in the presence
of fermion-bilinear Yukawa source terms. The fundamental fermion zero  
modes
$\chi_{i}$ and $\tilde{\chi}_{i}$, for $i=1,\ldots, N_{F}$, 
were constructed  in \refs{\CGTone,\Osborntwo}:
\eqn\fund{(\chi^{\alpha}_{i})^{\dot{\beta}}  \ = \
\bar{U}^{\dot{\beta}\alpha}_{\lambda}b_{\lambda k}f_{kl}{\cal K}_{li}
\ ,\quad
(\tilde{\chi}^{\alpha}_{i})^{\dot{\beta}}  \ = \
\bar{U}^{\dot{\beta}\alpha}_{\lambda}b_{\lambda
k}f_{kl}\tilde{\cal K}_{li}
}
with $\alpha$ a Weyl and $\dot\beta$ an $SU(2)$ color index. Using
\diffids a it is easily  checked that these are annihilated by
$\Dbarslash^{\dgamma}_{\ \alpha}.$ Note that each $\K_{ki}$ and  
$\Kt_{ki}$
is a Grassmann number rather than a Grassmann spinor; there is no
$SU(2)$ index. The normalization matrix of these modes is given by  
\Osborntwo
\eqn\orth{\int\, d^{4}x (\chi^{\alpha}_{i})^{\dot{\beta}}
(\tilde{\chi}_{\alpha j})_{\dot{\beta}}\ = \
\pi^{2}{\cal K}_{li}\tilde{\cal K}_{lj} 
}

Next we consider the adjoint Higgs bosons. In the presence of the
superpotential the Euler-Lagrange equation
\eulermatter{} for $A$ is unchanged; however Eq.~\Adageqn\ for
$A^\dagger$ now becomes
\eqn\newAdageqn{({\cal  
D}^{2}A^{\dagger})^{\dot{\gamma}}_{\,\,\,\dot{\alpha}}
\  = \
{1\over{2\sqrt{2}}}\,\sum_{i=1}^{N_{F}}(\chi_i^{\dot{\gamma}}
\tilde{\chi}^{}_{i\dot{\alpha}}+
\tilde{\chi}_i^{\dot{\gamma}}\chi^{}_{i\dot{\alpha}}
)\ ,}
displaying color and flavor  but suppressing Weyl indices. The 
solution of \newAdageqn\ is similar to, but simpler than, that of
\eulermatter{}. At the purely bosonic level, with all Grassmanns  
turned
off, $A$ and $A^\dagger$ must coincide, except for
 $\vhiggs\rightarrow\vhiggsbar.$ In contrast,
the fermion bilinear contributions to 
$A$ and to $A^{\dagger}$  in the path integral are to be treated as
independent.  
This bilinear contribution to $A^\dagger$ is straightforwardly 
obtained from \newAdageqn,
using the identity \diffids d, together with the manipulations  
described
in Sec.~7.2 of \dkmone. It has the form
\eqn\newblock{-i\Ubar^{\dalpha\alpha}\cdot
\pmatrix{0&\cdots&0 \cr
\vdots&\Ahyp\,\delta_\alpha^{\ \beta}&{}\cr 0&{}&{}}\cdot
U_{\beta\dbeta}\ ,}
where the $n\times n$ antisymmetric matrix $\Ahyp$ is defined as the
solution to the inhomogeneous linear equation
\eqn\Ahypdef{\bigL\cdot\Ahyp\ =\ \Lambdahyp\,.}
Here $\bigL$ is the ubiquitous linear matrix operator reviewed in   
Appendix A,
and the $n\times n$ antisymmetric matrix $\Lambdahyp$ is given by
\eqn\Lambdahypdef{(\Lambda_\hyp)_{k,l}\ =\ {i\sqrtwo\over16}\,
\sum_{i=1}^{N_F}\big(\K_{ki}\Kt_{li}+\Kt_{ki}\K_{li}\big)\ .}

\def\SNF{S^{N_F}_{\rm inst}}
\def\Szero{S^{0}_{\rm inst}}
Similarly, the squarks $q_i$ satisfy the leading-order Euler-Lagrange
equation
\eqn\squarkeq{\D^2 q_i\ =\ -i\sqrtwo\lambda\chi_i}
and likewise for $\tilde q_i$. Using Eq.~\diffids b together with
identities (7.9) and (C.3a) in \dkmone\ one easily derives
\eqn\squarkdef{q_i^\dbeta\ =\ \Ubar_\lambda^{\dbeta\beta}\cdot
\big(\,\delta_{\lambda0}\vhiggs_{i\beta}\,+\,\textstyle
{i\over2\sqrtwo}\M_{\lambda l\beta} f_{lk}\K_{ki}\,
\big)\ ,}
where $\vhiggs_{i}$ is the fundamental VEV of the $i$th hypermultiplet;
in the Coulomb branch all the $\vhiggs_i$ are zero.
All remaining adjoint and fundamental component fields of the  
superinstanton
in $N=2$ \susic\ QCD may be constructed by these methods.
 Fortunately, through judicious
use of integrations by parts together with the equations of motion,
the new expressions \fund, \newblock\ and \squarkdef\ are all that
are needed for our present goal of constructing the superinstanton
action, $\SNF$. By inspection of the component Lagrangian, one
sees that this action consists of a sum of five types
of terms:\hfil\break\indent(i) 
purely bosonic terms, \hfil\break\indent(ii) terms bilinear in the  
adjoint
fermion collective coordinates $\M$ and $\N$,\hfil\break\indent(iii)
 terms bilinear
in the fundamental fermion collective coordinates $\K$ and $\Kt,$ 
\hfil\break\indent(iv) 
fermion quadrilinear terms, consisting of one parameter drawn from 
each of $\M$, $\N$, $\K$ and $\Kt\,$, and  
finally\hfil\break\indent(v) 
the $N=2$
invariant hypermultiplet mass term.\hfil\break Let us consider each in  
turn.

The construction of (i),
(ii) and (iii) proceeds precisely along the lines discussed in detail
in Secs.~4.3 and 7.4 of \dkmone:  the relevant bits of the
component action are converted to
a surface term, and are given by
the coefficient of the $1/x^2$ fall-off of the total adjoint Higgs  
field,
including fermion bilinear contributions, as it approaches
its VEV. In this way one immediately
finds that the contributions (i) and (ii) to $\SNF$ are still given
by $\Szero,$ Eq.~\sinstfinal\ or \Scompact\ above. By identical  
arguments,
 (iii) is given by
\eqn\contribiii{-8\pi^2\Lambda_{lk}\A_{\hyp\, kl}\ ,}
where $\Lambda$ was defined in \Lambdabardef. 

More subtle is  the construction of the
fermion quadrilinear term (iv). Our calculation of this term proceeds  
in three
 steps, summarized 
as follows. \bf1\rm.~Show that $\bigL$ is self-adjoint, and use this
property to rewrite \contribiii\ as 
\eqn\newcontiii{-8\pi^2\Lambdahyp{}_{lk}\A'_{kl}\ }
where $\A'$ was defined in \Atotsum\ as the purely bosonic piece of
$\Atot$;
\bf2\rm.~Show that $\Lambdahyp$ is a \susic\ invariant; and finally
\bf3\rm.~Promote \newcontiii\ to a \susic\ invariant in the unique
way. Here are the details:

\bf1\rm. Let $\Omega$ and $\Omega'$ be two $n\times n$ antisymmetric
matrices that are scalar-valued (i.e., proportional to the identity
in the quaternionic space). Let
us define an inner product on the space of such
 matrices  in the naive way, by
\eqn\inprod{\langle\Omega'|\Omega\rangle\ =\ \Tr_n\,\Omega^{\prime  
T}\Omega\ .}
{}From the explicit expressions
in Appendix A, it is elementary to show that $\bigL$
is self-adjoint with respect to the above metric:
\eqn\selfadjoint{\langle\Omega'|\bigL\cdot\Omega\rangle
\ =\
\langle\Omega|\bigL\cdot\Omega'\rangle\ .}
The claimed equality between \contribiii\ and \newcontiii\ then follows
immediately from \selfadjoint\ and \Ahypdef\
together with the defining equation for $\A'$
(Eq.~(7.21) of \dkmone):
\eqn\twentyone{\bigL\cdot\A'\ =\ \Lambda\ .}

\bf2\rm. Next we show that each individual $\K_{ki}$ and $\Kt_{ki}$,
and hence $\Lambdahyp$, is a \susic\ invariant. As in Sec.~2 above,
we equate the active \susy\ transformation,\foot{To avoid clutter we
restrict ourselves here to the first \susy.}
\eqn\acthyp{\eqalign{\delta\,(\chi^{\alpha}_{i})^{\dot{\beta}} & = 
-i\sqrtwo\xibar_{1\dalpha}\Dbarslash^{\dalpha\alpha}q_i^\dbeta
\cr&=\ 
-\hf\xibar_{1\dalpha}\sigmabar^{n\dalpha\alpha}
\Ubar^{\dbeta\beta}\big(
\M_\beta f\partial_n(\Deltabar\Delta)
\,+\,
b\sigma_{n\beta\dgamma}f\Deltabar^{\dgamma\gamma}
\M_\gamma
\,\big) f\K_i\ ,}}
with the passive \susy\ transformation
\eqn\passhyp{\eqalign{\delta\,(\chi^{\alpha}_{i})^{\dot{\beta}} & = 
\delta\,\big(
\bar{U}^{\dot{\beta}\alpha}bf{\cal K}_{i}\big)
\cr&=\
\delta \Ubar^{\dbeta\alpha}bf\K_i+\Ubar^{\dbeta\alpha}b\delta f
\K_i+\Ubar^{\dbeta\alpha}bf\delta\K_i\ .}}
Remembering \Utrans, \entity\ and \susyalgebra{a},
 one finds that the first two terms on the
right-hand side of \passhyp\ equal the first two terms on the
right-hand side of \acthyp, respectively; this leaves
\eqn\deltaKzero{0\ = \delta\K_i\ =\ \delta\Kt_i\ ,}
as claimed.

\bf3\rm. Since $\delta\Lambdahyp=0,$  supersymmetrizing \newcontiii\
simply means promoting  
$\A'\rightarrow\Atot,$ as per Eqs.~\Atotsum\ and \susyalgebra{d}. Since
the difference between them consists of fermion bilinears, this step
introduces the promised fermion quadrilinears, and restores
\susy\ invariance.

Finally we turn to (v), the hypermultiplet mass term given in
\superp. In the Coulomb
branch this reduces to a mass term for the fundamental fermions only,  
up to
higher-order corrections in the coupling constant. From the  
normalization
condition \orth\ one derives
\eqn\massterm{S_{\rm mass}\ =\ \pi^{2}\sum_{i=1}^{N_{F}}m_{i}{\cal
K}_{li}\tilde{\cal K}_{li}\ .}
Putting these pieces together gives the general ADHM superinstanton
action for $N=2$ \susic\ QCD with gauge group $SU(2)$:
\eqn\SNFfinal{\SNF\ =\ \Szero\ -\
8\pi^2\Lambdahyp{}_{lk}\Atot_{kl}\ +\ S_{\rm mass}\ .}
This is the generalization to all $n$ of the 2-instanton action
presented recently in \refs{\dkmthree,\Aoyama}.

\newsec{General features of the super-multi-instanton action}

{}From the form of the action \SNFfinal\ we can immediately make several
observations of a general nature.

\bf1\rm. 
The self-adjointness property of $\bigL$ noted above allows
us to reexpress this action in a variety of equivalent ways. For  
instance
we can rewrite
\eqn\variety{8\pi^2\Lambdahyp{}_{lk}\Atot_{kl}\ =\
8\pi^2(\Lambda_{lk}+\Lambda_{f\,lk})\Ahyp{}_{kl}\ ,}
where the antisymmetric matrix $\Lambda_f$ introduced in \newmatdef\ 
is the Yukawa source term for $\A_f'$ (Eq.~(7.29) of \dkmone):
\eqn\twentynine{\bigL\cdot\A_f'\ =\ \Lambda_f\ .}

\def\Smass{S_{\rm mass}}
\def\muhyp{\mu_{\sst\rm hyp}}
\bf2\rm. 
Let us isolate $\Smass$ from the action \SNFfinal\ and assign it
to the $n\hbox{-}$instanton collective coordinate integration measure  
$d\muhyp$
for the fundamental fermions:
\eqn\muhypdef{\int d\muhyp\ =\
{1\over\pi^{2nN_F}}\int\prod_{i=1}^{N_F}d\K_{1i}\cdots d\K_{ni}\,
d\Kt_{1i}\cdots d\Kt_{ni}\,\exp(-\Smass)\ ,}
where the normalization constant in front has been read from Eq.~\orth. 
Consider this expression
in the chiral limit, $S_{\rm mass}=0.$
In this limit, for fixed flavor index $i,$ 
the Grassmann measure is obviously even or odd under the
discrete symmetry 
\eqn\discretion{\K_{li}\ \leftrightarrow\ \Kt_{li}\ ,}
 depending on whether $n$ itself is even or odd. On the other hand, the  
term
$-8\pi^2\Lambdahyp{}_{lk}\Atot_{kl}$ in the action \SNFfinal\
is always even under this symmetry, as follows from \Lambdahypdef. 
Therefore, for $N_F>0,$ 
 only the even-instanton sectors $n=0,2,4 \cdots$
can contribute in the chiral limit\foot{The
absence of a 1-instanton contribution is particularly easy
to see since $\Lambdahyp$  vanishes identically for $n=1$, and so the
$\K$ and $\Kt$ Grassmann integrations are unsaturated.}
(recall that  when $N_F=0$, all instanton numbers contribute).
This   selection rule was already noted 
by Seiberg and Witten in Sec.~3 of \SWtwo; it is not surprising to
find that it is built into the instanton calculus. 
It is violated when masses are turned on, since $\K_{li}\Kt_{li}$ is  
odd under
this symmetry.

\bf3\rm. In the even-instanton sectors (and in the
odd-instanton sectors as well when masses are turned on), 
 the number of exact fermionic modes, i.e.~those modes which do
not appear in the action, remains the same for all  $n$ and
for all $N_F$. Just as in  pure Yang-Mills theory \dkmone, 
the unbroken modes are the four adjoint fermionic modes
 associated with the four supersymmetry generators
that act nontrivially on the self-adjoint ADHM gauge field.\foot{This 
mode counting contrasts sharply with that of $N=1$ theories with
only fundamental Higgs bosons; see Appendix C for a brief discussion
of those types of models.} These are the 
\susic\ gaugino and Higgsino zero modes 
generated by $\xi_1Q_{1}$ and $\xi_2Q_{2}$, 
respectively. Explicitly,
they are given by Eq.~\lambdazm, with
\eqn\susysocalled{\M_\gamma\,=\,4\xi_{1\gamma}\,b  
\qquad\hbox{and}\qquad
\N_\gamma\,=\,4\xi_{2\gamma}\,b\ .}

\bf4\rm. 
Following the strategy originally established by
Affleck, Dine and Seiberg \ADSone, in the explicit calculations to  
follow
we will saturate these four unbroken modes
by suitable insertions of long-distance  fields. These
 are the components of the superinstanton that
are  parallel to the adjoint VEV and hence have power-law falloff; in 
comparison, the components orthogonal to the VEV decay exponentially as
$\exp(-M_W\, r),$ and can be ignored. As we saw 
 in pure Yang-Mills theory \dkmone, for any $n$ the structure of
these long-distance fields can be read off directly from 
the superinstanton action itself. In that theory, the  
long-distance 
antiHiggsino and antigaugino components satisfy \dkmone\
\eqna\LDfields
$$\eqalignno{\psibar_\dalpha(x)\ &=\ 
 i\sqrtwo\vhiggs^{-1}S_\higgs\,
\xi_1^\alpha {\rm S}_{\alpha\dalpha}(x,x_0) &\LDfields a\cr}$$
and 
$$\eqalignno{\lambdabar_\dalpha(x)\ &=\ 
 -i\sqrtwo\vhiggs^{-1}S_\higgs\,
\xi_2^\alpha {\rm S}_{\alpha\dalpha}(x,x_0)\ , &\LDfields b\cr}$$
respectively. Here $S_\higgs$ is the purely bosonic part of the
 superinstanton action,  $x_0$ is the position of the 
multi-instanton, 
and ${\rm S}(x,x_0)$ is the Weyl spinor propagator,
\eqn\spinprop{{\rm S}(x,x_0)\ =\ \delslash G(x,x_0)\ ,
\qquad
G(x,x_0)\ =\ {1\over4\pi^2(x-x_0)^2}\ .}
In \dkmtwo\ we found it helpful to rewrite \LDfields{} in a slightly  
different
way, as
\eqna\LDfieldstwo
$$\eqalignno{\psibar_\dalpha(x)\ &=\ 
 i\sqrtwo{\partial S_\inst^0\over\partial\vhiggs}\,
\xi_1^\alpha {\rm S}_{\alpha\dalpha}(x,x_0) &\LDfieldstwo a\cr}$$
and 
$$\eqalignno{\lambdabar_\dalpha(x)\ &=\ 
 -i\sqrtwo{\partial S_\inst^0\over\partial\vhiggs}\,
\xi_2^\alpha {\rm S}_{\alpha\dalpha}(x,x_0)\ , &\LDfieldstwo b\cr}$$
where in performing these derivatives
we  distinguish between $\vhiggs$ and $\vbarhiggs.$
Of course, in the pure Yang-Mills case, the expressions 
\LDfields{} and \LDfieldstwo{} 
are identical. This is because $S_\higgs$ is  linear in $\vhiggs$  
(bilinear
in $\vhiggs$ and $\vbarhiggs$), whereas the fermion-bilinear  
contribution
to $S_\inst^0$ depends only on $\vhiggsbar$ (see Eqs.~\sinstfinal\ and
\vevbcagain\ above). However, for $N_F>0$, this equality no longer  
holds,
since the new  term in the action,
$-8\pi^2\Lambdahyp{}_{lk}\Atot_{kl}$, depends on $\vhiggs$, not  
$\vbarhiggs$
(see Eqs.~\variety, \Lambdabardef\ and \vevbcagain). It 
turns out that the correct generalization to $N_F>0$ is
given, not by \LDfields{}, but  by the differential representation
\LDfieldstwo{} with $S_\inst^0$ replaced by $\SNF.$ As these  
long-distance
expressions
enter pervasively in the calculations below, we should make this
point especially clear; this is done in Appendix B.

Also needed below is the piece of the long-distance abelian
field strength $v_{mn}$ that is bilinear
in $\xi_1$ and $\xi_2$. The relevant expression is Eq.~(5.13) of
\dkmone\ which we likewise rewrite as a derivative:
\eqn\newbi{\sqrtwo\,{\partial S_\inst^0\over\partial\vhiggs}
\,\xi_1\sigma^{kl}\xi_2\,
G_{mn,kl}(x,x_0)\ ,}
where $G_{mn,kl}$ is the gauge-invariant propagator of $U(1)$ field  
strengths,
\eqn\Gmnkldef{G_{mn,kl}(x,x_0)\ =\ \big(
\,\eta_{nl}\partial_m\partial_k-\eta_{nk}\partial_m\partial_l
-\eta_{ml}\partial_n\partial_k+\eta_{mk}\partial_n\partial_l\,\big)G(x,x 
_0)\
 .}
As explained in Appendix B,
this expression, too, generalizes immediately to $N_F>0,$ with
the substitution
$S_\inst^0\rightarrow\SNF.$

\newsec{The prepotential}

In this Section we discuss some general features of the prepotential
$\F^{(N_F)}$ for $N=2$ \susic\ QCD. In Sec.~7.1, which is  restricted
to the cases $N_F<4$ unless otherwise stated, we review the
predictions of Seiberg and Witten \SWtwo\ (see also
Ref.~\refs{\Ohta-\Ito}). 
 Alternatively, 
in Sec.~7.2, we derive a formal expression, valid for $N_F\le4,$
for the prepotential in terms of the multi-instanton measure, extending
a result given in \dkmtwo\ to incorporate hypermultiplets.
Explicit numerical comparisons in the 1-instanton and 2-instanton
sectors will be given in Sec.~8 below.

\subsec{Seiberg-Witten predictions for the prepotential}

In $N=2$ SQCD with $N_{F}<4$ massless hypermultiplets in the
fundamental representation, the 
restrictions imposed by holomorphy, renormalization group invariance
and the anomaly imply that the
prepotential has the following expansion at weak coupling: 
\eqn\pre{
{\cal F}^{(N_{F})}({\rm v})\ =\ i\frac{(4-N_{F})}{8\pi}{\rm v}^{2}
\log{\left(\frac{C\,{\rm v}^{2}}{\Lambda^{2}_{N_{F}}}\right)}
-\frac{i}{\pi}\sum_{n=1}^{\infty}{\cal F}^{(N_{F})}_{n}\left(
\frac{\Lambda_{N_{F}}}{{\rm v}}\right)^{n(4-N_{F})}{\rm v}^{2} 
}
$\Lambda_{N_{F}}$ is the dynamically generated
scale of the theory\foot{The numerical values of the constants
${\cal F}^{(N_{F})}_{2k}$ depend on the definition of the scale 
$\Lambda_{N_{F}}$. In this paper, as in \dkmone, we adopt the 
$\Lambda$-parameter of the Pauli-Villars regularization scheme which
is appropriate for instanton calculations, and corresponds to
't~Hooft conventions for the collective coordinate measure
\tHooft. In this scheme
the renormalization group matching conditions are most
straightforward since the threshold factors are unity \FPone.
The $\Lambda$ of the Pauli-Villars 
scheme is related to the
$\Lambda$-parameter of \refs{\SWone,\SWtwo} as 
$4\Lambda^{4-N_{F}}_{N_{F}}=(\Lambda^{\rm SW}_{N_{F}})^{4-N_{F}}$
for $0 \le N_{F} < 4$.} and $C$ is a numerical constant. 
The logarithm comes from  the classical result
combined with one-loop perturbation theory while the remaining 
terms correspond to an infinite series of  
instanton corrections. As discussed above, for $N_F>0$
the  discrete symmetry \discretion\ ensures that
only even numbers of instantons contribute: hence 
${\cal F}^{(N_{F})}_{2k+1}=0$. Each non-zero 
coefficient ${\cal F}^{(N_{F})}_{2k}$ is a pure number characterizing
the leading semiclassical contribution of instantons of topological
charge $2k$.  In Sec.~8 below,
we carry out an explicit two-instanton computation of
${\cal F}^{(N_{F})}_{2}$ for $N_F\le4$.
 For the special case of $N_{F}=4$ massless hypermultiplets
the $\beta$-function  is zero and we 
expect the following expression for the prepotential:
\eqn\prefour{
{\cal F}^{(4)}({\rm v})\ =\ \frac{1}{4} \ \tau_{\rm cl}
\ {\rm v}^{2} \
- \ \frac{i}{\pi}\sum_{n=2,4,6\ldots}\,{\cal F}^{(4)}_{n}\,q^n
\ {\rm v}^{2} \ ,\qquad
q\,=\,\ e^{i  \pi \tau_{\rm cl}}
} 
Furthermore,  Seiberg and Witten 
propose that the massless $N_{F}=4$ theory is classically exact,
Eq.~\counterpart,
which implies $ {\cal F}^{(4)}_{n} = 0$ for all $n$.
Instead, in Sec.~8 we will obtain a non-zero value for ${\cal F}^{(4)}_{2}$.

The general description of the low-energy theory,
Eqs.~\pre-\prefour, is modified in
two ways by the introduction of masses for the
hypermultiplets. First, as noted earlier,
the mass terms explicitly break the discrete
symmetry \discretion; hence the contribution of odd numbers of
instantons becomes non-zero. Second, each term in the instanton
expansion will itself be a polynomial in the dimensionless ratios
$m_{i}/{\rm v}$. As we will see below, these polynomials 
can be obtained from the exact solution of the low energy theory  
proposed
by Seiberg and Witten.

For each $N_{F}$, the exact behavior of the low-energy
theory is characterized by an elliptic curve of the form 
\eqn\curve{y^{2}\ =\ x^{3}+Bx^{2}+Cx+D\ \equiv\ (x-e_1)(x-e_2)(x-e_3)\  
.}
For $N_{F}<4$, where the theory is asymptotically free, 
the coefficients $B$, $C$ and $D$ (and hence the roots $e_i$)
are functions of the modulus $u$,
the dynamical scale $\Lambda_{N_{F}}$, and the masses $m_{i}$. 
The exact solution for the VEV 
${\rm v}$ and its dual ${\rm v}_{D}$ as functions of the modulus $u$
is given in terms of the periods of the elliptic curve 
\curve. In the region of parameter space where the
roots are real and $e_{1}\geq e_{2}\geq e_{3}$, we have the explicit
formulae (in the conventions of \FPone),
\eqn\solution{
\frac{\partial {\rm v}}{\partial u}\ =\  
\frac{{\rm c}K(k)}{\sqrt{e_{1}-e_{3}}} \ ,
 \qquad  \frac{\partial{\rm  v}_{D}}{\partial u}\ =\ 
\frac{i{\rm c}K'(k)}{\sqrt{e_{1}-e_{3}}}\ .}
Here $K$ and $K'$ are elliptic functions of the first kind,  
$k=\sqrt{(e_{2}-e_{3})/(e_{1}-e_{3})}$, and ${\rm c}$ is a numerical
constant fixed by demanding the asymptotic behavior 
${\rm v}=\sqrt{2u}+\ldots$ in the weakly-coupled regime of large $u$. 
The second derivative of the prepotential is then given as
\eqn\pretwo{\frac{\partial^{2} {\cal F}}{\partial {\rm v}^{2}}\ =\  
{1\over2}
\frac{\partial{\rm v}_{D}}{\partial {\rm v}}\ =\ \frac{iK'(k)}{2K(k)}\  
.}
This equation, together with \solution, 
determines the prepotential corresponding to a
particular elliptic curve up to irrelevant constants of integration.

With the definition of the $\Lambda$-parameter given above, 
the elliptic curve for the $N_{F}=0$ theory is simply
\eqn\ymcurve{y^{2}_{(0)} \ =\  x^{2}(x-u)+ \Lambda^{4}_{0}x\ .}
For $0< N_{F}<4$, the curves can be written in terms of the following  
set of
$SO(2N_{F})$-invariant polynomials in the masses $m_{i}\,$:
\eqn\inv{\eqalign{
M^{(N_{F})}_{0} \ &= \ 1\cr
M^{(N_{F})}_{1} \ &=\ \sum_{i=1}^{N_{F}} m_{i}^{2}\cr
M^{(N_{F})}_{2} \ &=\ \sum_{i<j}^{N_{F}} m^{2}_{i}m^{2}_{j} \cr
\vdots\ \qquad         &\qquad{} \ \vdots \cr
M^{(N_{F})}_{N_{F}}\  &=\ \prod_{j=1}^{N_{F}} m_{j}^{2}    }}
The curves are then given by
\eqn\curves{
y^{2}_{(N_{F})} \ =\  
x^{2}(x-u)\,+\,\sqrt{M^{(N_{F})}_{N_{F}}}\Lambda^{4-N_{F}}_{N_{F}}x
\,-\,\quarter\Lambda^{2(4-N_{F})}_{N_{F}}
\sum_{\delta=0}^{N_{F}-1}M^{(N_{F})}_{\delta}
(x-u)^{N_{F}-1-\delta}}
Given these explicit forms, it is straightforward to expand equations
\solution\ and \pretwo\ as a power series in 
$\Lambda^{4-N_{F}}_{N_{F}}$ and extract the first few terms in the
instanton expansion of the prepotential. However, as we will see
below, several features of the expansion can be deduced without 
further calculation.

Let us order the masses so that they satisfy $m_{N_{F}}\geq
m_{N_{F}-1}\geq\ldots \geq m_{1}\geq0\,.$ 
An important restriction on the form of the elliptic curves comes from 
the scaling limit \hbox{$m_{N_{F}}\rightarrow \infty$},
$\Lambda_{N_{F}}\rightarrow 0$ with
$m_{N_{F}}\Lambda^{4-N_{F}}_{N_{F}}$ held fixed. In this limit 
one of the flavors becomes infinitely massive and decouples, leaving
an effective theory described by $N=2$ SQCD with $N_{F}-1$ flavors.    
In the chosen regularization scheme, 
the $\Lambda$-parameters for different numbers of flavors are related  
as 
\eqn\mrelated{m_{N_{F}}\Lambda^{4-N_{F}}_{N_{F}} \  =\
\Lambda^{5-N_{F}}_{N_{F}-1}\ .}
It is easy to check that the curves \curves\ have the required
property $y^{2}_{(N_{F})} \rightarrow y^{2}_{(N_{F}-1)}$ in the
decoupling limit. This property is then inherited by the prepotential  
itself:  
\eqn\decouple{
{\cal F}^{(N_{F})}({\rm v}; \{ m_{i} \},\Lambda_{N_{F}})\  
\longrightarrow \
{\cal F}^{(N_{F}-1)}({\rm v}; \{ m_{i},\, i<N_F\},\Lambda_{N_{F}-1})
\ .}
Moreover, this relation must hold order by order in the instanton
expansion.

The single-instanton factor $\Lambda^{4-N_{F}}_{N_{F}}$  appears in
\curves\ multiplied by the product $m_{1}m_{2}\cdots m_{N_{F}}$. 
Obviously this is the only term in \curves\ capable of generating
odd powers of $\Lambda_{N_F}^{4-N_F}$. It follows that
the odd terms in the instanton expansion of
the  prepotential vanish unless all of the masses are non-zero, as
expected from the discrete symmetry \discretion. 

The one-instanton contribution has the form 
\eqn\oneinst{ {\cal F}^{(N_{F})}({\rm v};
\{m_{i}\},\Lambda_{N_{F}})\Big|_{n=1}\ =\ -\frac{i}{\pi}
\frac{\Lambda_{N_{F}}^{4-N_{F}}}{{\rm v}^{2}}{\cal F}^{(0)}_{1}\,
\prod_{j=1}^{N_{F}}m_{j}\ .}
This clearly obeys the decoupling relation \decouple.
The numerical coefficient ${\cal F}^{(0)}_{1}$ can be extracted from 
the instanton expansion of the prepotential of the $N_{F}=0$  
theory
\refs{\FPone,\dkmone}: ${\cal F}^{(0)}_{1}=1/2$ .

The remaining terms in \curves\ are proportional to 
$\Lambda^{2(4-N_{F})}_{N_{F}}$ and can therefore be thought of as a 
two-instanton effect. In particular, note that the term proportional to 
$(x-u)^{N_{F}-1}$ remains non-zero in the massless limit. 
Every term of order $\Lambda^{2(4-N_{F})}_{N_{F}}$ which can be
formed from the coefficients of the elliptic curve is proportional to
one of the polynomials $M^{(N_{F})}_{\delta}$ defined above. 
Hence, by dimensional analysis, the 
two-instanton contribution to the prepotential must have the form 
\eqn\expo{
 {\cal F}^{(N_{F})}({\rm v};\left\{ m_{i} \right\},
\Lambda_{N_{F}})\Big|_{n=2}
\ =\  -\frac{i}{\pi}{\rm v}^{2}\left(\frac{\Lambda_{N_{F}}}
{{\rm v}}\right)^{2(4-N_{F})}
\sum_{\delta =0}^{N_{F}} {f}^{(N_{F})}_{\delta}\left(\frac{
M_{\delta}^{(N_{F})}}{{\rm v}^{2\delta}}\right)}

This expression may be constrained further by considering various  
limits
of the masses. In the chiral limit, $m_{i}\rightarrow 0$, we
recover the coefficients of \pre; this forces
$f^{(N_{F})}_{0}={\cal F}^{(N_{F})}_{2}$. 
In the opposite limit $m_{N_F}\rightarrow\infty,$ 
the decoupling relation \decouple\
implies that the numerical coefficients
$f^{(N_{F})}_{\delta}$ are not independent, but obey
\eqn\rone{f^{(N_{F})}_{\delta}\ =\ f^{(N_{F}-1)}_{\delta -1}\ =\  
\ldots\ =\ 
f^{(N_{F}-\delta)}_{0}\ .}
It follows that the constant $f^{(N_{F})}_{\delta}$ is equal to the 
coefficient ${\cal F}^{(N_{F}-\delta)}_{2}$ of the massless case. 
An explicit calculation using \solution-\pretwo\ yields the values 
${\cal F}^{(0)}_{2}=5/2^{4}$, ${\cal F}^{(1)}_{2}=-3/2^{5}$ 
and ${\cal F}^{(2)}_{2}=1/2^{6}$. 
The coefficient ${\cal F}^{(3)}_{2}$ 
corresponds to an additive constant in the prepotential which does not
contribute to the low-energy effective Lagrangian and is irrelevant
for our purposes.

In component form, for any number of flavors,
the low-energy effective Lagrangian is defined
in terms of the prepotential as follows (the superscript
$\scriptstyle\rm SD$ stand for ``self-dual''):
\eqn\Leffcomp{\eqalign{\Leff\ =\ {1\over4\pi}\,\Im\Big[
&-\F''(A)\Big(\partial_mA^\dagger\partial^mA+i\psi\delslash\psibar
+i\lambda\delslash\lambdabar+\hf(\vsd_{mn})^2\,\Big)
\cr&+{\textstyle{1\over\sqrtwo}}\F'''(A)\lambda\sigma^{mn}\psi  
v_{mn}+\quarter
\F''''(A)\psi^2\lambda^2\,\Big]\ .}}
As usual we  ignore auxiliary fields as they are subleading in the 
coupling
constant. The last three terms in  the above Lagrangian yield  
non-vanishing
tree-level contributions to the following three Green's functions  
\dkmone:
\eqna\foureff
$$\eqalignno{\langle v_{mn}(x_{1})v_{kl}(x_{2})\rangle 
\ &=\
\frac{1}{16\pi i}\frac{\partial^{2}{\cal F}}
{\partial {\rm v}^{2}}
\int\, d^{4}x_{0}\,
{\rm tr}_{2}\sigma^{pq}\sigma^{rs}\cr
&\qquad\times
 G_{mn,pq}(x_{1},x_{0})G_{kl,rs}(x_{2},x_{0}) &\foureff a \cr
\langle v_{mn}(x_{1}) \bar{\lambda}_{\dot{\alpha}}(x_{2})
\bar{\psi}_{\dot{\beta}}(x_{3})\rangle\  &=\
\frac{1}{8\sqrt{2}\pi i}\,\frac{\partial^{3}{\cal F}}
{\partial {\rm v}^{3}}
\int\, d^{4}x_{0}\,
\sigma^{kl\alpha\beta}G_{mn,kl}(x_{1},x_{0})\cr
&\qquad\times
{\rm S}_{\alpha\dot{\alpha}}(x_{2},x_{0}) 
{\rm S}_{\beta\dot{\beta}}(x_{3},x_{0}) &\foureff b \cr
\langle \bar{\lambda}_{\dot{\alpha}}(x_{1})
\bar{\lambda}_{\dot{\beta}}(x_{2})
\bar{\psi}_{\dot{\gamma}}
(x_{3})\bar{\psi}_{\dot{\delta}}(x_{4})\rangle \ &=\
\frac{1}{8\pi i}\frac{\partial^{4}{\cal F}}
{\partial {\rm v}^{4}}
\int\, d^{4}x_{0}\,
\epsilon^{\alpha\beta}
{\rm S}_{\alpha\dot{\alpha}}(x_{1},x_{0})&\foureff c \cr
&\qquad\times
{\rm S}_{\beta\dot{\beta}}(x_{2},x_{0}) 
 \epsilon^{\gamma\delta}
{\rm S}_{\gamma\dot{\gamma}}(x_{3},x_{0})
{\rm S}_{\delta\dot{\delta}}(x_{4},x_{0})
}$$
where the Weyl and field-strength propagators ${\rm S}(x,x_{0})$ 
 and $G_{mn,kl}(x,x_{0})$  were defined in \spinprop\ and \Gmnkldef\  
above.
We now discuss how these correlation functions may be calculated from 
first principles, using instanton methods.

\subsec{The prepotential in the instanton approach}

\def\LD{{\sst\rm LD}}
Our strategy for determining ${\cal F}$ is to calculate the leading
semiclassical contributions to the  Green functions \foureff{}
in the large distance limit. The first step is to replace each of the  
fields
$\psibar,$ $\lambdabar$  and $v_{mn}$ with the
long-distance ``tail'' of the corresponding component of the  
superinstanton.
These expressions, which we denote
$\psibar^\LD,$ $\lambdabar^\LD$  and $v_{mn}^\LD$,
 were given above in Eqs.~\LDfieldstwo a, \LDfieldstwo b, and
\newbi, respectively, with the substitution $S_\inst^0\rightarrow\SNF$.

One also needs the superinstanton measure. 
In the $n$-instanton sector, the integration runs over $8n$
bosonic and $8n+2nN_{F}$ fermionic collective coordinates, which
we denote generically as $X_i$ and $\chi_i$, respectively. At a purely
formal level, the measure for this integration can be expressed as
\eqn\measure{
\int\, d\mu^{(N_{F})}_{n}\ =\ \frac{1}{{\cal S}_{n}}
\int\, \left(\prod_{i=1}^{8n}dX_{i}\prod_{j=1}^{8n+2nN_{F}}d\chi_{j} 
\right)(J_{{\rm bose}}/J_{{\rm fermi}})^{1/2}\exp\left(-\SNF(n)\right)}
Here $J_{{\rm bose}}$ and $J_{{\rm fermi}}$ are the collective
coordinate Jacobians for the bosonic and fermionic parameters
respectively and ${\cal S}_{n}$ is a symmetry factor. 

As reviewed in \dkmone, it is only possible to solve the ADHM
constraints and find an explicit formula for the measure for 
$n< 3$. However, for the following, it suffices to know that the
only dependence on the VEV ${\rm v}$ in \measure\ is that of the
action $\SNF$ which is separately linear in both 
 ${\rm v}$ and $\bar{\rm v}$. In addition, as discussed in Sec.~6,
 we know that
all fermionic zero modes are lifted
by the action except for the four supersymmetric zero  modes  
\susysocalled\
parametrized by
$\xi_{1\alpha}$ and $\xi_{2\alpha}$. It is convenient to separate
out from the measure these unbroken modes together with
 their bosonic partner, the translational
degrees of freedom, $x_{0}\,$:
\eqn\mutildedef{\int\,d\mu^{(N_{F})}_{n}\ =\  
\int\,d^{4}x_{0}\,d^{2}\xi_1\,
d^{2}\xi_2\int\, d\tilde{\mu}^{(N_{F})}_{n}\ .}
We will refer to $d\tilde{\mu}^{(N_{F})}_{n}$ as the ``reduced  
measure.''

Putting the pieces together, one finds for the $n$-instanton  
contribution 
to the Green's function \foureff c:
\eqn\res{ \langle \bar{\lambda}_{\dot{\alpha}}(x_{1})
\bar{\lambda}_{\dot{\beta}}(x_{2})
\bar{\psi}_{\dot{\gamma}}
(x_{3})\bar{\psi}_{\dot{\delta}}(x_{4})\rangle 
 \ =\   \int d\mu^{(N_{F})}_{n}  
\bar{\lambda}^{\scriptscriptstyle\rm LD}_{\dot{\alpha}}(x_{1})
\bar{\lambda}^{\scriptscriptstyle\rm LD}_{\dot{\beta}}(x_{2})
\bar{\psi}^{\scriptscriptstyle\rm LD}_{\dot{\gamma}}
(x_{3})\bar{\psi}^{\scriptscriptstyle\rm LD}_
{\dot{\delta}}(x_{4}) \ ,}
with similar expressions for the other two Green's functions. Following 
\dkmtwo, we substitute the expressions \LDfieldstwo{}-\newbi\ into
the right-hand side, and perform the
trivial integration over $\xi_{1\alpha}$ and 
$\xi_{2\alpha}$.  This leaves:
\eqna\greenies
$$\eqalignno{
\langle v_{mn}(x_{1})v_{kl}(x_{2})\rangle 
\ &=\
\frac{1}{2}\frac{\partial^{2}}
{\partial {\rm v}^{2}}
\int\,d\tilde{\mu}^{(N_{F})}_{n}\, \int\, 
d^{4}x_{0}\,{\rm tr}_{2}\sigma^{pq}\sigma^{rs}
\cr
&\qquad\times
 G_{mn,pq}(x_{1},x_{0})
G_{kl,rs}(x_{2},x_{0})  
&\greenies a\cr
\langle v_{mn}(x_{1}) \bar{\lambda}_{\dot{\alpha}}(x_{2})
\bar{\psi}_{\dot{\beta}}(x_{3})\rangle  \ &=\
\frac{1}{\sqrt{2}}\frac{\partial^{3}}
{\partial {\rm v}^{3}}
 \int\,d\tilde{\mu}^{(N_{F})}_{n}\, \int\, d^{4}x_{0}\,
\sigma^{kl\alpha\beta}&\greenies b
\cr
&\qquad\times
G_{mn,kl}(x_{1},x_{0})
{\rm S}_{\alpha\dot{\alpha}}(x_{2},x_{0}) 
{\rm S}_{\beta\dot{\beta}}(x_{3},x_{0})  \cr
\langle \bar{\lambda}_{\dot{\alpha}}(x_{1})
\bar{\lambda}_{\dot{\beta}}(x_{2})
\bar{\psi}_{\dot{\gamma}}
(x_{3})\bar{\psi}_{\dot{\delta}}(x_{4})\rangle\ &=\
\frac{\partial^{4}}{\partial {\rm v}^{4}}
\int\,d\tilde{\mu}^{(N_{F})}_{n}\, \int\, d^{4}x_{0}\,
\epsilon^{\alpha\beta}
{\rm S}_{\alpha\dot{\alpha}}(x_{1},x_{0})
\cr
&\times
{\rm S}_{\beta\dot{\beta}}(x_{2},x_{0}) 
\epsilon^{\gamma\delta}
{\rm S}_{\gamma\dot{\gamma}}(x_{3},x_{0})
{\rm S}_{\delta\dot{\delta}}(x_{4},x_{0})&\greenies c 
}$$
The linearity of $\SNF$ in ${\rm v}$  has allowed us to pull the 
$\vhiggs$ differentiation outside the collective coordinate integral.
Comparing the semiclassical expressions \greenies{} with their exact
counterparts \foureff{}, we deduce 
\eqn\ndeq{{\cal F}^{(N_{F})}({\rm v};\{m_{i}\},
\Lambda_{N_{F}})\Big|_{n\rm-inst}\ =\  8\pi i\,\int\, 
d\tilde{\mu}^{(N_{F})}_{n}\,+\,{ A}{\rm v}\,+\,{ B}\ ,}
where ${A}$ and ${B}$ are undetermined constants of
integration. As these constants do not contribute to
the low-energy effective Lagrangian, we are free to set $A=B=0$ for
convenience.

Equation \ndeq\ is the desired expression for the prepotential as a  
formal
integral over the superinstanton moduli. It is the obvious  
generalization
to $N_F>0$ of the analogous SYM formula (21) in \dkmtwo. Note that all  
memory
of the long-distance field insertions has disappeared from this  
equation.
In hindsight, these insertions were merely a convenient bookkeeping
device for extracting the appropriate derivatives of $\cal F$
dictated by the low-energy  Lagrangian \Leffcomp. Henceforth we
will drop all reference to the ``tail'' of the superinstanton, and
focus directly on the concise expression \ndeq.

\newsec{Explicit calculations for $n=1$ and $n=2$}

In this section we will use the action derived above to calculate the
one- and two-instanton contributions to the prepotential for 
$N_{F}\leq 4$. In addition we include non-zero masses 
for the hypermultiplets. This section supplies  additional
details for the calculation of $\langle\psibar\psibar\lambdabar\lambdabar
\rangle\,\sim\,\F''''$ presented in our recent Letter \dkmthree.
See also \Aoyama\ for a related calculation of a different
quantity which is not simply given by a derivative of $\F$. 
By focusing on $\F$ itself using Eq.~\ndeq, we also extract
information about the special case $N_F=4.$

\subsec{The 1-instanton contribution}

In ADHM language, the bosonic and fermionic parameters of a single
$N=2$ superinstanton are contained in three $2\times 1$ matrices of
unconstrained parameters:
\eqn\oneI{a \,=\, \left({w \atop X} \right) \ ,\qquad
{\cal M}_{\gamma}\, =\,\left(
{\mu_{\gamma} \atop M_{\gamma}}\right)\ ,\qquad
{\cal N}_{\gamma}\,=\,\left(
{ \nu_{\gamma} \atop N_{\gamma}}\right)\ .}
In addition there are $2N_{F}$ Grassmann variables ${\cal K}_{i}$ and 
$\tilde{\cal K}_{i}$ which parametrize the fundamental zero modes
\fund.  The reduced measure \mutildedef\ is given  by
\eqn\oneImeasure{\int \, d\tilde{\mu}^{(N_{F})}_{1}
  \ =\   \frac{2^{7}\Lambda^{4-N_{F}}_{N_{F}}}
{\pi^{4+2N_{F}}}\int\,d^{4}wd^{2}\mu d^{2}\nu\times 
\prod_{i=1}^{N_{F}} d{\cal K}_{i}d\tilde{\cal K}_{i}
\exp\big(-\tilde\SNF(n=1)\big)\ , }
where the single-superinstanton action is easily read from the general
expression \SNFfinal:\foot{We place a tilde over the action to indicate
that the Maxwell piece $8\pi^2 n/g^2$ has been subtracted out.}
\eqn\oneIaction{\tilde\SNF(1)\   = \ 16\pi^{2}|w|^{2}
|\vhiggsa|^2
+
4\sqrt{2}\pi^{2}\mu \barvhiggsa \nu+\pi^{2}\sum_{i=1}^{N_{F}}\, m_{i} 
{\cal K}_{i}\tilde{\cal K}_{i}\ .}
(For the special case
$N_{F}=4$, where the $\beta$-function
vanishes, the factor $\Lambda_{N_{F}}^{4-N_{F}}$ should simply 
be replaced by
$q$ from Eq.~\qdeff.)
Notice that the only dependence on 
${\cal K}_{i}$ and $\tilde{\cal K}_{i}$ comes from the mass
term $S_\mass$ (this is because $\Lambdahyp$ vanishes identically for  
$n=1$,
as do $\Lambda$ and $\Lambda_f$). The corresponding Grassmann 
integrations can only be saturated by
 bringing down $N_{F}$ powers of $S_\mass$;
 as expected from the discrete symmetry \discretion, the result is
nonzero only when all the  $m_{i}$ are nonzero.
 The remaining integration is identical to the
case of $N=2$ SYM theory \refs{\FPone,\dkmone}. Using
\ndeq\ one finds after a simple calculation:
\eqn\oneIprepot{ {\cal F}^{(N_{F})}({\rm v};
\{m_{i}\},\Lambda_{N_{F}})\Big|_{n=1}\ =\ -\frac{i}{\pi}
\frac{\Lambda_{N_{F}}^{4-N_{F}}}{{\rm v}^{2}}{\cal F}^{(0)}_{1}\,
\prod_{j=1}^{N_{F}}m_{j}\ ,}
where ${\cal F}^{(0)}_{1}=1/2$,
 in agreement with the Seiberg-Witten prediction.
\subsec{The $2$-instanton contribution}

Next  we consider the 2-instanton contribution to the
prepotential. The notation and the various changes of integration
variables in the present calculation closely parallel
the simpler case of $N=2$ SYM theory worked out in
detail in Sec.~8 of \dkmone. Because the calculation for $N_F>0$
is so similar, we will chiefly stress those points where they differ.

The parameters of the $n=2$ ADHM superinstanton are
contained in the following $3\times 2$ matrices:
\eqna\twoinstmats
$$\eqalignno{
a\ &=\ \pmatrix{w_1&w_2\cr x_0+a_3&a_1\cr a_1&x_0-a_3}\ \  
,&\twoinstmats a
\cr
\M_\gamma\ &=\ \pmatrix{\mu_{1\gamma}&\mu_{2\gamma}\cr
4\xi_{1\gamma}+\M_{3\gamma}&\M_{1\gamma}\cr
\M_{1\gamma}&4\xi_{1\gamma}-\M_{3\gamma}}\ ,&\twoinstmats b \cr
\N_\gamma\ &=\ \pmatrix{\nu_{1\gamma}&\nu_{2\gamma}\cr
4\xi_{2\gamma}+\N_{3\gamma}&\N_{1\gamma}\cr
\N_{1\gamma}&4\xi_{2\gamma}-\N_{3\gamma}}\ .&\twoinstmats c}$$
In addition, there are now $4N_{F}$ fundamental zero modes \fund\  
parametrized
by the Grassmann numbers ${\cal K}_{li}$ and $\tilde{\cal
K}_{li}$ with $l=1,2$. We also define the following frequently  
occurring
combinations of these collective coordinates:
\eqn\useful{\eqalign{
L \ &=\ |w_{1}|^{2}+|w_{2}|^{2}
\ \cr
H \ &=\ |w_{1}|^{2}+|w_{2}|^{2} + 4|a_{1}|^{2}+4|a_{3}|^{2} \cr
\Omega\ &=\ w_{1}\bar{w}_{2}-w_{2}\bar{w}_{1} \cr 
\omega\ &=\ \bar{w}_{2}\vhiggsa w_{1}-\bar{w}_{1}\vhiggsa w_{2}\ 
=\ \hf\trtwo\Omega\vhiggsa \ =\ -\Lambda_{1,2}\cr
Y \ &=\ \mu_{1}\nu_{2}-\nu_{1}\mu_{2}+2{\cal M}_{3}{\cal N}_{1}
-2{\cal N}_{3}{\cal M}_{1} \ =\ 2\sqrtwo\,(\Lambda_f)_{1,2} \cr
Z \ &=\ \sum_{i=1}^{N_{F}}{\cal K}_{ki}\epsilon_{kl}\tilde{\cal K}_{li} 
\ =\ -8\sqrtwo i\,(\Lambdahyp)_{1,2}}}

{}From \SNFfinal\ we write down the following expression for the
2-instanton action \hbox{$\tilde\SNF(n=2)\,$}:
\eqn\SNFtwoI{\tilde\SNF(2)\ =\ \tilde\Szero(2)\ -\
8\pi^2\Lambdahyp{}_{lk}\Atot_{kl}\ +\ S_{\rm mass}\ .}
The hypermultiplet mass term
 $S_\mass$ is given in \massterm\ above, whereas
$\tilde\Szero(2)$ was evaluated in \dkmone, and is
concisely expressed in terms of the quantities \useful:
\eqn\oldanswer{\tilde\Szero(2)\ =\ 
16\pi^2L|\vhiggsa|^2\,+\,4\sqrtwo\pi^2\mu_k\barvhiggsa\nu_k\,-\,
{16\pi^2\omegabar\over H}\big(\,\omega-{Y\over2\sqrtwo}\,\big)\ .}
The remaining term in the action,
$-8\pi^2\Lambdahyp{}_{lk}\Atot_{kl}$, is easily extracted from  
\Lambdahypdef, 
together with the defining equation for $\Atot,$ namely (A.1) below.
The linear operator $\bigL$ that enters that expression
is a $\big[\hf n(n-1)\times\hf n(n-1)\big]$-dimensional map from the
space of $n\times n$ antisymmetric matrices onto itself. When $n=2$ 
this space is one-dimensional, and $\bigL$ reduces
 to ordinary multiplication by the quantity $H$.  From 
\Lambdabardef, \newmatdef\ and \useful, one therefore
finds for this term:
\eqn\twoInew{-{i\sqrtwo\pi^2Z\over H}\,
\big(\,\omega-{Y\over2\sqrtwo}\,\big)\ .}
Note that \twoInew\ may be absorbed  into the SYM action
\oldanswer\ with the simple substitution 
\eqn\twoIsub{\omegabar\ \longrightarrow\ \omegabar\,+\,iZ/8\sqrtwo\ .}

Given the action, the next step is the construction of the 2-instanton
measure. We begin by eliminating the redundant degrees of freedom from
 \twoinstmats{}. A convenient resolution of the 
 ADHM constraints \crucial a and \zmcons a is to
 eliminate the off-diagonal elements $a_{1}$, ${\cal M}_{1\gamma}$, 
and ${\cal N}_{1\gamma}$, as follows:
\eqna\aonedef
$$\eqalignno{ a_1\ &=\ {1\over4|a_3|^2}\,a_3(\wbar_2w_1-\wbar_1w_2)\ ,
&\aonedef a
\cr
\M_1\ &=\ {1\over2|a_3|^2}\,a_3\,\big(\,2\abar_1\M_3+\wbar_2\mu_1
-\wbar_1\mu_2\,\big)\ ,&\aonedef b}$$
and
$$\eqalignno{\N_1\ &=\  
{1\over2|a_3|^2}\,a_3\,\big(\,2\abar_1\N_3+\wbar_2\nu_1
-\wbar_1\nu_2\,\big)\ .&\aonedef c\cr}$$
The remaining degrees of freedom are unconstrained, and appear
as integration  variables in the measure.
 It is helpful to factor the reduced measure
$d\tilde\mu_2^{N_F}$
 into three parts
$d\tilde{\mu}_{b}$, 
$d\tilde{\mu}_{f}$
 and $d\muhyp$ 
corresponding to the bosonic, adjoint fermionic, and fundamental
fermionic parameters, respectively:
\eqn\divone{\int d\tilde{\mu}^{(N_{F})}_{2}  \ =\  
\int\,d\muhyp \,d\tilde{\mu}_{b}\,d\tilde{\mu}_{f} \ .}
Here $d\muhyp$ was defined in Eq.~\muhypdef\ above,
\eqn\muhypagain{\int d\muhyp\ =\
{1\over\pi^{4N_F}}\int\prod_{i=1}^{N_F}d\K_{1i} d\K_{2i}\,
d\Kt_{1i} d\Kt_{2i}\,\exp(-\Smass)\ ,}
 while 
$d\tilde{\mu}_{b}$ may be read off from \dkmone, subject to the  
replacement
\twoIsub, and to the appropriate redefinition of the dynamically
generated scale parameter:
\eqn\defmua{\eqalign{\int d\tilde{\mu}_{b}\ &=\ \frac{2^{10}
\Lambda^{2(4-N_{F})}_{N_{F}}}{\pi^8{\cal S}_{2}}
\int\, d^{4}a_{3}d^{4}w_{1}d^{4}w_{2} \frac{\big|
|a_{3}|^{2}-|a_{1}|^{2} \big|}{H}\cr&\qquad\qquad\times\
\exp\left(-16\pi^2\big[\,L|\vhiggsa|^2\,-\,
{\omega\over H}\,\big(\omegabar+iZ/8\sqrtwo\big)\,\big]\right)\ .}}
The symmetry factor ${\cal S}_{2}=16$ is  associated with a
discrete redundancy in the chosen parametrization \aonedef a
of the two instanton
solution \refs{\dkmone,\Oone}.
 For the special case
$N_{F}=4$, where the $\beta$-function
vanishes, the factor $\Lambda_{N_{F}}^{2(4-N_{F})}$ should simply 
be replaced by
$q^2$.
 The third piece of the measure comprises
the remaining terms in the action:
\eqn\defmub{\eqalign{\int d\tilde{\mu}_{f} \ &=\
\int\, d^{2}{\cal M}_{3}d^{2}\mu_{1}d^{2}\mu_{2} 
d^{2}{\cal N}_{3}d^{2}\nu_{1}d^{2}\nu_{2} \cr&\qquad\qquad\times\
\exp\left(
-4\sqrtwo\pi^2\big[\,\mu_k\barvhiggsa\nu_k+
{Y\over H}\,\big(\omegabar+iZ/8\sqrtwo\big)\,\big]\right)\ .}}

Performing the Grassmann integration over the 
parameters of the adjoint zero modes is a straightforward exercise; one
finds
\eqn\lift{\eqalign{
\int d\tilde{\mu}_{f}\  &=\
-\left(\frac{16\sqrt{2}\pi^{6}(\bar{\omega}+iZ/8\sqrtwo)}
{|a_{3}|^{2}H}
\right)^{2}\Big[\sixteenth\vbarhiggs^4
|\Omega |^{2} 
+\frac{L}{2H}
\vbarhiggs^2
(\bar{\omega}+iZ/8\sqrtwo)\bar{\omega}
 \cr
& \qquad{} \qquad{} \qquad{}   +\ {1\over H^2}(\bar{\omega}+
iZ/8\sqrtwo)^{2}
\left(\quarter\vbarhiggs^2
(L^{2}-|\Omega|^{2})+\bar{\omega}^{2}\right)\Big]
}}
This is the generalization to $N_F>0$ of the Yukawa determinant
given in Eq.~(8.13) \hbox{of \dkmone.}
The next step is to integrate over the fundamental fermionic
coordinates using the following identity:
\eqn\littlest{
\int\,d\muhyp\, G(Z)\ =\ \sum_{k=0}^{N_{F}}
\frac{M^{({N_{F}})}_{N_{F}-k}}{\pi^{4k}}
\frac{\partial^{2k}G}{\partial Z^{2k}}\Big|_{Z=0}\ , }
where the $M^{(N_{F})}_{\delta}$ are the polynomials defined in \inv\
above.  This is the only new feature involved for $N_F>0.$

Finally we turn to the remaining integration over the bosonic moduli.
Following \dkmone, it is convenient to change variables in the
bosonic measure from $\{a_{3},w_{1},w_{2}\}$ to the new 
set  $\{H,L,\Omega\}$.  The relevant formulae are:
\eqn\covone{\int_{-\infty}^\infty d^4a_3\ 
{\big|\,|a_3|^2-|a_1|^2\,\big|\over |a_3|^4}\ \longrightarrow\
{\pi^2\over2}\,\int_{L+2|\Omega|}^\infty\ dH}
and 
\eqn\covtwo{\int_{-\infty}^\infty\,d^4w_1\,d^4w_2\ 
\longrightarrow\
{\pi^3\over8}\int_0^\infty dL\ \int_{|\Omega|\le L}d^3\Omega\ .}
The numerator and denominator in the left-hand side of \covone\ are
supplied by \defmua\ and \lift, respectively.
In addition we introduce rescaled variables $\Omega=L\Omega'$, $H=LH'$,
and $\omega=L\omega'$. Following \dkmone\ we now carry out the
trivial integration over $L$. Thanks to \littlest, at this stage
in the calculation the RG decoupling property \expo-\rone\ is
manifest; this is another example of a general feature of the
hyperelliptic curves being built into the instanton calculus.

Finally we switch to
 spherical polar coordinates, 
\eqn\polarcoor{d^3\Omega'\ \longrightarrow\ 2\pi\int_{-1}^1d(\cos
\theta)\int_0^1|\Omega'|^2d|\Omega'|\ ,}
where the polar angle is defined by $|\omega'|=|\Omega'||\vhiggsa|
\cos\theta=\hf|\Omega'||\vhiggs|\cos\theta.$ This leaves an
ordinary 3-dimensional scaleless integral over the remaining
variables $H,$ $\cos\theta$ and $|\Omega'|$ which is the precise
analog of Eq.~(8.19) in \dkmone. Performing this  elementary
integral with the help of a standard symbolic manipulation routine
gives:
%
%
\eqn\exactvalues{{\cal F}^{(0)}_{2}\,=\,5/2^{4}\ ,\quad
{\cal F}^{(1)}_{2}\,=\,-3/2^{5}\ ,\quad
{\cal F}^{(2)}_{2}\,=\,1/2^{6}\ ,\quad
{\cal F}^{(3)}_{2}\,
=\,-5/(2^{7}3^{3})\ .}
These values of ${\cal F}^{(N_F)}_{2}$ with $N_F=0,1,2$ agree with
the predictions extracted from \refs{\SWone,\SWtwo}. 
 ${\cal F}^{(3)}_{2}$
 corresponds to a constant shift in the prepotential, which does
not affect the low-energy Lagrangian \Leffcomp. These numbers
are the input parameters for Eqs.~\pre\ and \expo.
Finally, for the conformally invariant case
$N_{F}=4$ we likewise find a non-vanishing result,
\eqn\nonexactvalue{{\cal F}^{(4)}_{2}\,=\,7/(2^{8}3^{5})\ , }
which is associated with the series \prefour. 
This implies Eq.~\tauresult\ which
is in contradiction with the classical exactness \counterpart\
proposed in \SWtwo.

$$\scriptscriptstyle*************************$$

We acknowledge useful discussions with
S. Elitzur,
T. Harano,
M. Matone,
M. Peskin,
M. Sato, and 
A. Yung.
The work of ND was supported in part by a PPARC Advanced Research
Fellowship; both ND and
 VVK were supported in part by the Nuffield 
Foundation; MM was supported by the Department of Energy. VVK thanks
the CERN theory group for hospitality.

\appendix{A}{Supersymmetric invariance of $\Atot$}

The purpose of this Appendix is to demonstrate that the $n\times n$
antisymmetric matrix $\Atot$ is in fact a \susic\ invariant, $\delta
\Atot=0,$ given the transformation laws \susyalgebra{a\hbox{-}c}.
Our starting point is the defining equation for $\Atot$:
\eqn\thirtysomething{\bigL\cdot\Atot\ =\ \Lambdatot\ ,}
where $\Lambdatot=\Lambda+\Lambda_f$, while $\Atot=\A'+\A'_f$ is the
quantity under examination. Note that this equation is the sum of
Eqs.~\twentyone\ and \twentynine\ that we used in the text.
$\bigL$ is a linear operator that maps the space of $n\times n$ 
 scalar-valued antisymmetric matrices onto itself. Explicitly,
if $\Omega$ is such a matrix, then $\bigL$ is defined as \dkmone
\eqn\bigLreally{\bigL\cdot\Omega\ =\ 
\hf\{\,\Omega\,,\,W\,\}\,-\,\hf\trtwo\big(
[\,\abar'\,,\,\Omega\,]a'-\abar'[\,a'\,,\,\Omega\,]\big)}
where $a'$ was defined in \bcanonical\ and \symcond\ above, and
$W$ is the symmetric scalar-valued $n\times n$ matrix
$W_{kl}\,=\,\wbar_kw_l+\wbar_lw_k\, .$

Applying a general $N=2$ \susy\ variation to \thirtysomething\ gives
\eqn\varyAtot{\bigL\cdot\delta\Atot\ =\ \delta\Lambdatot\,-\,
\delta\bigL\cdot\Atot\ .}
Since $\bigL$ is generically invertible, it suffices to show that
the right-hand side vanishes. To minimize clutter we restrict the  
variation
to $\xibar_2\Qbar_2$, as the calculation with  
$\Qbar_2\rightleftharpoons
\Qbar_1$ proceeds identically, while the claim for $Q_1$ and $Q_2$
is a trivial consequence of \zmcons b. We define the $n$-vectors
\def\nuv{\vec\nu}
\def\wv{\vec w}
\def\wbarv{\vec{\wbar}}
\eqn\nvecdef{\nuv=(\nu_1,\cdots,\nu_n)\ ,\quad
\wv=(w_1,\cdots,w_n)\ ,\quad
\wbarv=(\wbar_1,\cdots,\wbar_n)\ .}
Starting with the most complicated term on the right-hand
side of \varyAtot, one  finds
\eqn\rhsfirst{\eqalign{
&\delta\Lambda_f\ =\ 
\Big(\nuv^T\vhiggsa\wv\xibar_2-\xibar_2\wbarv^T\vhiggsa\nuv\Big)-
\Big(\nuv^T\wv\cdot\Atot\xibar_2-\xibar_2\Atot\cdot\wbarv^T\nuv\Big)
\cr&\quad
+\Big(\N^{\prime T}[\Atot,a']\xibar_2+\xibar_2[\Atot,\abar']\N'\Big)}}
using \susyalgebra{b}. The first term in big
parentheses on the right-hand side precisely cancels $\delta\Lambda$;
 the second term in big parentheses groups together
with $-\hf\{\,\Atot\,,\,\delta W\,\}$ to give 
$-\hf\big[
\nuv^T\wv\xibar_2+\xibar_2\wbarv^T\nuv\,,\,\Atot\big]\,$;
the third term in big parentheses combines with the remaining
terms on the right-hand side of \varyAtot\ to give 
$-\hf\big[\N^{\prime T}a'\xibar_2+\xibar_2\abar'\N'\,,\,\Atot\big]\,.$
 Since
$\wbarv^T\nuv+\abar'\N'=\abar\N$
these two commutators add to
\eqn\addto{-{1\over2}\big[\N^Ta\xibar_2+\xibar_2\abar\N\,,\,\Atot\big]}
which  vanishes by virtue of \zmcons a. \sl QED \rm

\appendix{B}{Long-distance fields in $N=2$ \susic\ QCD.}

In this Appendix we justify the differential expressions
\LDfieldstwo{}-\newbi\ for the long-distance ``tail'' of the superinstanton,
specifically in the background of the \susic\ adjoint zero modes 
\susysocalled. 
As stated in Sec.~6, for the case $N_F=0$ these expressions are equivalent
rewritings of the formulae in \dkmone\ (e.g., Eq.~\LDfields{} above), but for
$N_F>0$ they differ by $\xi\K\Kt$ Grassmann trilinears.

{}From the explicit component Lagrangian including the superpotential
\superp, together with the identities \diffids{}, one can of course
solve the Euler-Lagrange equations
for these fields explicitly. But it is easier to exploit the
$N=2$ \susy\ algebra itself to generate these solutions automatically. 
As in Sec.~4 of \dkmone,  we use a construction due to
Refs.~\NSVZ\ and \FSone: One starts with a ``reference''
superinstanton $\Psi^\zero$ comprising a convenient initial choice of
component fields (bosonic and fermionic, adjoint and fundamental),
and generates the desired configuration by acting on it with the appropriate
symmetry generators.

For present purposes $\Psi^\zero$ is specified as follows. In
the hypermultiplets $Q_i$ and $\tilde Q_i,$ fill only the fundamental
fermion zero modes \fund; the remaining fundamental fields are
turned off. In contrast, in the adjoint sector, fill only the bosonic
components initially. Thus the gauge field is the usual ADHM configuration,
while the Higgs bosons $\Abar$ and $A$ satisfy, respectively,
Eq.~\newAdageqn, and the homogeneous equation
$\D^2 A=0.$ All antifermions are initially zero.
Note that the components of $\Psi^\zero$ correctly obey the
leading-order coupled Euler-Lagrange equations. 

Next one acts on $\Psi^\zero$ infinitesimally with the $N=2$ generators
$\sum_{i=1,2}\,\xi_iQ_i$. This action generates the desired
\susy\ modes \susysocalled\ in the adjoint gaugino and Higgsino
components. At the same time it produces nonzero  antigaugino
and 
antiHiggsino fields that automatically satisfy their respective Euler-Lagrange
equations in this background.
The $N=2$ algebra gives:
\eqn\antifsdef{\lambdabar\ =\ i\sqrtwo\xitwo\Dslash\Abar\ ,
\qquad\psibar\ =\ -i\sqrtwo\xione\Dslash\Abar\ ,}
where $\Abar$ obeys \newAdageqn\ as stated above. In particular $\Abar$
has not only
a pure bosonic part but a part bilinear in $\K\,\Kt$ as well, which
implies trilinear Grassmann contributions to $\lambdabar$ 
\hbox{and $\psibar.$}

Now think about the behavior of \antifsdef\  far away from
 the center of the 
multi-instanton. On the one hand (in generalized singular gauge \dkmone), 
the $x$-dependence and the tensor structure of the right-hand
side approaches  a spinor propagator \spinprop, up to ${\cal O}(x^{-5})$
corrections.
On the other hand, from  the divergence theorem, we actually know the
 coefficient of the leading $1/x^3$ fall-off as well. Specifically, this
coefficient can be equated to certain terms in
 the superinstanton action. This follows from an integration
by parts in the Higgs kinetic energy term in the component Lagrangian,
together with the Higgs equation of motion to cancel  the Yukawa term; 
see Secs.~4.3 and 7.4 of \dkmone\ for details. In the case of $N=2$
SYM theory the entire superinstanton action $S_\inst^0$ may be read
off from the residue in this way.
 In the models at hand, with $N_F>0,$ this particular surface integral
only accounts for two of the five separate pieces of $S_\inst^{N_F}$ listed
in Sec.~5 above, namely those labeled (i) and (iii);
these are precisely the pieces proportional to $\vhiggs.$ 
Combining these two observations, about the tensor structure and
about the residue, gives  Eq.~\LDfieldstwo{}; for $N_F>0$
one needs to substitute $S_\inst^0\rightarrow S_\inst^{N_F}$ as stated.

In the same way, the desired $\xi_1\xi_2$ bilinear piece of the
field strength $v_{mn}$ may be generated from $\Psi^\zero$ by acting
with $\exp(\xi_2Q_2)\exp(\xi_1Q_1)$ and keeping the cross-term.
Under $Q_1$ one has $\delta v_{mn}=\xi_1\sigma_{[\,n}\D_{m\,]}\lambdabar$,
which is followed by the replacement \antifsdef\ under the action of
$Q_2.$ The remaining steps in the argument proceed as before
(see Sec.~5 of \dkmone), and are
left to the reader.

\appendix{C}{$N=1$ theories with fundamental Higgs bosons}

In this Appendix we touch on certain basic features of
the $N=1$ \susic\ theories of the type
considered long ago in \refs{\NSVZ,\ADSone,\FSone}, in which
all Higgs bosons live in the fundamental representation
of $SU(2)$. We can easily construct the superinstanton action by the  
methods
of \dkmone, as follows. The two relevant terms of the component
Lagrangian, the Higgs kinetic energy and the Yukawa interaction, are
turned into a surface term with an integration by parts in the former
together with the Euler-Lagrange equation for the fundamental scalar,
$q$. As per the divergence theorem, the action
may then be extracted from the $1/x^3$ fall-off of $\D_{\sst\perp} q,$  
where
 the normal covariant derivative
$\D_{\sst\perp}$ is defined as $(x^m/\sqrt{|x|^2})\,\D_m\,$. The  
generic
form of $q$ including fermion bilinear contributions was given in 
Eq.~\squarkdef\ above:
\eqn\squarkagain{q^\dbeta\ =\ \Ubar_\lambda^{\dbeta\beta}\cdot
\big(\,\delta_{\lambda0}\vhiggs_{\beta}\,+\,\textstyle
{i\over2\sqrtwo}\M_{\lambda l\beta} f_{lk}\K_{k}\,
\big)\ ,}
ignoring flavor indices from now on. As in the text the $\K$ are the  
Grassmann 
parameters associated with the fundamental fermion zero modes \fund.

Using \diffids a together with the asymptotics of the various ADHM
quantities listed in Sec.~6.2 of \dkmone, one easily derives
\eqn\funasym{
\D_{\sst\perp} q^\dbeta\quad
 \buildrel |x|\rightarrow\infty\over\longrightarrow
\quad {\sigmabar_0^{\dbeta\beta}\over|x|^3}\,
\sum_{k=1}^n
\,\big(|w_k|^2\vhiggs_\beta\ -\ {i\over\sqrtwo}\,\mu_{k\beta}\K_k\big)}
and hence
\eqn\funaction{S_\inst\ \propto\ 
\sum_{k=1}^n
\,\big(|w_k|^2|\vhiggs|^2\ -\ {i\over\sqrtwo}\,
\vbarhiggs^\beta\mu_{k\beta}\K_k\big)}
using the notation of Eq.~\bcanonical.

This \susic\ multi-instanton 
action (originally derived by Yung \oldyung\ by different means)  
differs from
 that of the $N=2$ theory discussed herein, in two important ways.
First, it has the form of a disconnected sum of $n$ single instantons;  
in
these coordinates there is no interaction between them.
Second, the only gaugino modes that are lifted are those associated
with the top-row elements $\mu^\gamma_k$ of the collective coordinate  
matrix
$\M^\gamma$. This leaves $2n$ unlifted modes (after one implements the 
constraints \zmcons{}), which are naturally associated with the
diagonal entries of the $n\times n$ submatrix $\M_\gamma'$. This 
counting contrasts
sharply with the $N=2$ theory in which the number of unlifted modes
is independent of the winding number.

Saturating these modes with anti-Higgsinos as per 
Affleck, Dine and Seiberg \ADSone, one therefore
needs $\langle\psibar(x_1)\psibar(x_2)\rangle$ in the 1-instanton
sector, and in general,
 $\langle\psibar(x_1)\cdots\psibar(x_{2n})\rangle$ in the $n$-instanton
sector---unlike the $N=2$ theory, the sectors of
different topological number do not interfere with one another.
For completeness we write down the generic form of these antifermions,
which satisfy the inhomogeneous equation
\eqn\funl{(\Dslash_{\alpha\dalpha}\psibar^\dalpha)^\dbeta
\ =\ c(\lambda_\alpha)^\dbeta{}_\dgamma\,q^{\dagger\dgamma}}
where $c$ is a normalization constant. Using \diffids a once again,
one easily finds
\eqn\funm{(\psibar^\dalpha)^\dbeta\ =\
-(c/2)\,\Ubar^{\dbeta\beta}\M_\beta f\wbar^{\dalpha\gamma}
\vbarhiggs_\gamma\ .}
Here $\dbeta$ and $\dalpha$ are color and Weyl indices, respectively;
also \funm\ is only valid when the top row of $\M$ (i.e., the lifted
modes) consists entirely of zeroes.

\vfil
\listrefs
\bye